\documentclass[11pt]{article}
\usepackage{amsmath}
\usepackage{geometry}                
\usepackage{graphicx}
\usepackage{amssymb}
\usepackage{epstopdf}
\title{Lectures on the holographic duality of gauge fields and strings}
\author{Gordon W. Semenoff\\~\\Department of Physics and Astronomy, University of British Columbia, \\
6224 Agricultural Road, Vancouver, British Columbia, Canada V6T 1Z1}
\date{}                                           

\begin{document}
\maketitle
\vskip .9cm
\centerline{
Lectures presented at the Les Houches 2016 Summer School} 
\vskip .4cm
\centerline{``INTEGRABILITY: FROM STATISTICAL SYSTEMS TO GAUGE THEORY'' }
\vskip .9cm

\centerline{\large\bf Outline}

\begin{enumerate}

\item{}Prologue

\item{}Preamble about gauge fields 

\item{}The Wilson Loop

\item{}The large N expansion

\item{}D branes, black branes and  Maldacena's conjecture

\item{} ${\mathcal N}=4$ superconformal Yang-Mills theory

\item{}Holographic Wilson loops

\item{}Epilogue




\end{enumerate}

 \section{Prologue}
 
 This is a summary of lecture notes that I have presented as a pedagogical review of some of the basics of the holographic duality between string
theory and gauge field theory. I presented these lectures to a mixed audience of students and early
career researchers who were all theoretical physicists but with diverse backgrounds. I have made no
attempt to be complete, or to give a comprehensive review of the subject or even a comprehensive introduction.   
There are plenty of reviews \cite{Petersen:1999zh}
 \cite{Aharony:1999ti}  \cite{DiVecchia:1999yr}
\cite{DiVecchia:1999yr} 
\cite{Semenoff:2002kk} \cite{Bigazzi:2003ui}
\cite{Maldacena:2003nj} \cite{McGreevy:2009xe}
\cite{Polchinski:2010hw}
\cite{Hubeny:2011hd}
 and even a few excellent books \cite{Ramallo:2013bua}
\cite{Ammon:2015wua}
 \cite{zaanen} \cite{Natsuume}   \cite{nastase}
 \cite{sean} available and they more than  adequately fill the gap.  
What I have prepared is a discussion of some aspects off the AdS/CFT correspondence as I best understand them, with the
hope that it will help students of the school where the lectures are presented 
appreciate the more advanced courses on more specialized topics which will come later.

\section{Preamble about gauge fields}

In this section  we shall begin by
sketching some of the basics of Yang-Mills theory \cite{Yang:1954ek}.  I assume that students have seen such an introduction
already.  What I have written here 
 is intended to define some of the ideas and to fix some of the notation and conventions for 
the material that will come afterward.  

\subsection{Basics}

 Throughout these lectures, we will use the natural system of units where the physical constants $\hbar=1$ and $c=1$. 
We will  consider quantum field theories in either Minkowski space or Euclidean space.  In four space-time dimensions, 
 the Cartesian coordinates of Minkowski space are  the time, $x^0$,  and the three space dimensions, $x^1,x^2,x^3$. 
 They combine to form the position four-vector, denoted by $x^\mu$.  
 Lorentz indices are lowered and raised by the Minkowski metric,
\begin{align}\label{minkowski_metric}
\eta_{\mu\nu}= \left[ \begin{matrix} -1 & 0 & 0 & 0 \cr 0 & 1 & 0 & 0 \cr
0 & 0 & 1 & 0\cr 0 & 0 & 0 & 1\cr \end{matrix}\right]
\end{align} 
and its inverse, $\eta^{\nu\lambda}$ (so that $ \eta_{\mu\nu}\eta^{\nu\lambda}
\equiv \sum_{\nu} \eta_{\mu\nu}\eta^{\nu\lambda}=\delta_\mu^{~\lambda}$). In this expression and hereafter, we use
the Einstein summation convention where repeated up and down indices are assumed to be summed.  In Euclidean space, the
metric is simply the unit matrix,  $\delta_{\mu\nu}$. An incremental translation $dx^\mu$ typically has an up-index and 
a derivative by a four-vector has down indices, $\partial_\mu = \frac{\partial}{\partial x^\mu}$. These indices can be lowered and
raised by the metric, $dx_\mu\equiv \eta_{\mu\nu}dx^\nu$ or $\partial^\mu = \eta^{\mu\nu}\partial_\nu$. 

The basic dynamical variable of Yang-Mills theory is a connection field, $$A_\mu(x)$$ with a lower Lorentz
index, so that the combination $A\equiv A_\mu(x)dx^\mu$ is a one-form. 
At the same time as being a one-form, the connection takes values in the Lie algebra of the gauge group.  We will practically
exclusively use the Lie group $U(N)$ for the gauge group. 
In that case, for each value of the index $\mu$
and for each value of coordinates, $x$, $A_\mu(x)$ is a 
Hermitian matrix.   It obeys $A_\mu^\dagger(x)=A_\mu(x)$ where
\begin{align}
[A^\dagger_\mu(x)]_{ab} = [A^*_\mu(x)]_{ba}
\end{align}
Sometimes, we will also be interested in the Lie algegra for the Lie group  $SU(N)$.  In that case
$A_\mu(x)$ are traceless as well as Hermitian matrices. 
It is possible and straightforward to generalize our  discussion
 to other Lie algebras if it is needed. 

The connection field  is used to form covariant derivatives.  Covariant derivatives are designed to act on wave-functions. 
The wave-function of a quark is a space-time dependent object with $N$ complex components, 
which we shall denote by \begin{align}\label{psi}\Psi(x)=\left[\begin{matrix}\psi_1(x)\cr \psi_2(x)\cr \ldots\cr\psi_N(x)\cr\end{matrix}\right]\end{align}
 A gauge transformation maps it to 
\begin{align}\label{gt}
\Psi(x)\to g(x)\Psi(x)
\end{align}
where, for each value of the space-time coordinates, $x$,  $g(x)$ is an $N\times N$ unitary
matrix, that is, an $N\times N$ matrix which obeys $g^\dagger(x)g(x)=1$ and $g(x)g^\dagger(x)=1$.  
The covariant derivative of the wave-function is defined as 
\begin{align}
D_\mu\Psi(x)~\equiv ~\left(\partial_\mu -iA_\mu(x)\right)\Psi(x)
\end{align}
The purpose of a covariant derivative to to define a derivative in such a way that it is compatible with
gauge transformations. Accordingly, the covariant derivative of the wave-function 
 must gauge transform in the same way as the wave-function, that is,
\begin{align}\label{gauge_transform}
D_\mu \Psi(x) ~\to ~g(x)D_\mu\Psi(x)
\end{align}
This fixes the gauge transformation of the connection so that, the gauge transformations are
\begin{align}
A_\mu(x)~&\to~g(x) A_\mu(x) g^{\dagger}(x) -i\partial_\mu g(x)g^\dagger(x) \label{gt2}\\
\Psi(x)~&\to~ g(x)\Psi(x)
\label{gt3}
\end{align}

A basic notion of geometry is that of parallel transport.  It is a way of transporting a vector along a curve in such a way that it remains
``parallel''  to its original orientation. 
Here, the wave-function $\Psi(x)$  in (\ref{psi}) is the $N$ component vector with complex valued components 
and a change in orientation of this vector is  the multiplication
by a $N\times N$ unitary matrix, as in (\ref{gt3}).   
We can think of these gauge transformations (\ref{gt2}) and  (\ref{gt3})  as  the analog 
of  general coordinate transformations in Riemanian geometry.   
The data which defines what is meant by ``parallel''  is stored
 in the connection $A_\mu(x)$.  The result of parallel transport of the wave-function $\Psi(x_1)$ along a curve $C$ to the point $x_2$ is
  $
 \Psi(x_2)=
 U_C(x_2,x_1)\Psi(x_1)
$ where   $ U_C(x_2,x_1)$ is a unitary matrix which we will construct in the paragraphs below.
 
 Consider the parametric representation of a curve, $C$,  in space-time given by four functions $x^\mu(\tau)$ of the curve parameter $\tau$.
 As $\tau$ runs over its range, the functions $x^\mu(\tau)$ trace the curve $C$ in space-time.  For the purposes of the following argument, $C$
 is an open curve (that is, a curve with distinct beginning and end-points).

  The wave-function is constant along the curve $C$ if, at each point $\tau$ of that curve, 
\begin{align}\label{covconst}
\dot x^\mu(\tau)D_\mu\Psi(x(\tau))=0
\end{align}
 We have used the notation $\dot x^\mu(\tau)=\frac{d}{d\tau}x^\mu(\tau)$. 
Equation (\ref{covconst}) can be integrated to show that $\Psi(x)$ is constant along the curve $C$ if, 
for any two points $x_1^\mu=x^\mu(\tau_1)$ and $x_2^\mu
 =x^\mu(\tau_2)$ lying on $C$, 
 $$
 \Psi(x_2)=
 U_C(x_2,x_1)\Psi(x_1)
$$ 
where  $U_C(x_2,x_1)$ is  the path ordered exponential of the line integral of the connection field along $C$, 
\begin{align}
U_C(x_2,x_1)&={\mathcal P}e^{i\int_{\tau_1}^{\tau_2}d\tau \dot x^\mu(\tau)A_\mu(x(\tau))}\\
&\equiv \sum_{n=0^\infty}\frac{i^n}{n!}\int_{\tau_1}^{\tau_2} d\tau_1\dot x_{\mu_1}(\tau_1)\ldots\int_{\tau_1}^{\tau_2} d\tau_n\dot x_{\mu_n}(\tau_n)
{\mathcal P}A_{\mu_1}(x(\tau_1))\ldots  A_{\mu_n}(x(\tau_n))
\end{align}  
The symbol ${\mathcal P}$ orders the matrices so that the those with later arguments are to the left of those with earlier arguments. 
$U_C(x_2,x_1)$ is itself a unitary matrix which tells us how the wave-function $\Psi(x)$ is re-oriented as it is parallel transported along $C$. 
Moreover,
$$
U_C^{-1}(x_2,x_1)=U_C^\dagger(x_2,x_1)=U_C(x_1,x_2)
$$
 Also, under the gauge transformation (\ref{gt2}) and (\ref{gt3}), 
\begin{align}
U_C(x_2,x_1)~\to ~g(x_2) U_C(x_2,x_1)g^{-1}(x_1)
\end{align}
so the parallel transported wave-function has the correct gauge transformation property,
 \begin{align}
 U_C(x_2,x_1)\Psi(x_1)
\to g(x_2)
 U_C(x_2,x_1)\Psi(x_1)~~{\rm when}~~\Psi(x_1)\to g(x_1)\Psi(x_1)~,\nonumber \\~A_\mu(x)~\to~g(x) A_\mu(x) g^{\dagger}(x) -i\partial_\mu g(x)g^\dagger(x)
\label{gt4}\end{align}
 Generally, parallel transport depends on the path, $C$.  A measure of its path dependence is  called the curvature of the connection,
defined as a commutator of covariant derivatives, 
\begin{align}\label{curvature}
F_{\mu\nu}(x)=i \left[ D_\mu,D_\nu\right] = \partial_\mu A_\nu-\partial_\nu A_\mu-i\left[ A_\mu, A_\nu\right]
\end{align}
Each component of this curvature tensor is itself an $N\times N$ Hermitian matrix. Under a gauge transformation,  
\begin{align}
F_{\mu\nu}(x)~\to~g(x)F_{\mu\nu}(x)g^\dagger(x)
\end{align}
In any region of space-time where $F_{\mu\nu}\neq 0$ the parallel transport of a wave-function depends on the curve
along which it is transported.   $F_{\mu\nu}$ cannot be any arbitrary set of six Hermitian matrices.  It is constrained by the
fact that it is derived from a connection, as in equation (\ref{curvature}).  The fact that the curvature is a commutator of covariant 
derivatices plus the Jacobi identity for commutators
$$
\left[ D_\mu \left[ D_\nu, D_\lambda\right]\right]+\left[ D_\mu \left[ D_\lambda, D_\mu\right]\right]+
\left[ D_\lambda \left[ D_\mu, D_\nu\right]\right]=0
$$
require that $F_{\mu\nu}(x)$ satisfies the Bianchi identity, 
\begin{align}\label{bianchi}
D_\mu F_{\nu\lambda}(x) +  D_\nu F_{\lambda\mu}(x) + D_\lambda F_{\mu\nu}(x) =0
\end{align}
Note that, in the covariant derivative of   $F_{\mu\nu}(x)$, the connection appears
in a commutator,
$$
D_\mu F_{\nu\lambda}(x) \equiv \partial_\mu F_{\nu\lambda}(x)-i\left[ A_\mu(x),   F_{\nu\lambda}(x)\right] 
$$

The quark wave-function that we have considered here is an $N$-component complex vector and it transforms in
the fundamental representation of the
gauge group, that is, like $\Psi(x)\to g(x)\Psi(x)$. 
There are other representations that are of interest to us. 
 One which will appear a lot in the
following is the ``adjoint representation''.  In that case, the wave-function is  an $N\times N$ complex matrix 
and its gauge transformation is
\begin{align}
\psi(x) \to g(x)\psi(x) g^{-1}(x)
\end{align}
 The covariant derivative acts on such a wave-function as
$$
D_\mu \Psi(x) = \partial_\mu \Psi(x) -i\left[ A_\mu(x) , \Psi(x)\right]
$$
Under parallel transport along $C$, it obeys
\begin{align}
\Psi(x_1)~\to ~ U_C(x_2,x_1)\Psi(x_1) U^\dagger_C(x_2,x_1) =  U_C(x_2,x_1)\Psi(x_1) U_C (x_1,x_2)
\end{align}
This consideration can easily be
generalized to wave-functions which transform in higher representations of the gauge group. 

\subsection{Yang-Mills theory}

In Yang-Mills theory, the connection $A_\mu(x)$  becomes a dynamical variable. The connection itself is a redundant description of this variable, more properly said,
the gauge orbit of a connection is the dynamical variable.  A gauge orbit is an equivalence class of connections, where the equivalence relation is $A\sim A'$ if $A$ and $A'$
are related by a gauge transformation.  The principle of gauge invariance requires that all physical quantities in Yang-Mills theory should be gauge invariant.   This principle
is particularly important when the Yang-Mills theory is quantized since the logical and mathematical consistency of the theory depends on it. 

The dynamics of a classical or quantum mechanical theory must be fixed by specifying the
equations of motion for the dynamical variables.  
The equations of motion can be obtained from an action by using a variational principle.  The usual action is one which
is quadratic in the curvature, so that the dynamics tend to favour field configurations with vanishing curvature,  
\footnote{
We will at times assume that the four dimensional
spacetime is Euclidean, rather than Lorentzian.  In the Euclidean case, $-{\rm Tr}F_{\mu\nu}F^{\mu\nu}$ is replaced 
by ${\rm Tr}F^{\mu\nu}F_{\mu\nu}$ which we sometimes write as ${\rm Tr}F_{\mu\nu}F_{\mu\nu}$. 
The factor of $\frac{1}{2}$ in the Lagrangian density is
conventional.   There is another way of writing gauge fields by expanding the connection in 
a complete basis for $N\times N$ Hermitian matrices,  $T^a$, $a=1,2,...,N^2$
\begin{align}
A_\mu(x) =\sum_{a=1}^{N^2} T^a A^a_\mu(x)
\end{align}
Also 
\begin{align}
F_{\mu\nu}(x) =\sum_{a=1}^{N^2} T^a F^a_{\mu\nu}(x)
\end{align}
Typically, the basis matrices have a commutation relation 
\begin{align}
[T^a,T^b]=if^{abc}T^c
\end{align}
and they are normalized so that ${\rm Tr}T^aT^b = \frac {1}{2}\delta^{ab}$.
With this notation, the Lagrangian density has
the   conventional normalization 
\begin{align}
{\mathcal L}_{\rm YM} =- \frac{1}{4g_{\rm YM}^2}\sum_{a=1}^{N^2}F^a_{\mu\nu}F^{a\mu\nu}
\end{align}
with $F^a_{\mu\nu}=\partial_\mu A^a_\nu-\partial_\nu A^a_\mu-f^{abc}A_\mu^b A_\nu^c$.}
\begin{align}\label{Yang_Mills_action}
S_{\rm YM} = \int dx~ {\mathcal L}_{\rm YM}~~,~~~~~~~~ {\mathcal L}_{\rm YM}= -\frac{1}{2g_{\rm YM}^2}{\rm Tr}\left(F_{\mu\nu}F^{\mu\nu}\right)
\end{align}

Here, the integration is over space-time.   The symbol Tr means taking the trace of the matrix 
\begin{align}\label{trace}{\rm Tr}M\equiv\sum_{a=1}^N M_{aa}
\end{align} 
Matrices appearing in a trace can be permuted cyclically,
${\rm Tr}ABC = {\rm Tr}BCA = {\rm Tr}CAB$. One can use this cyclicity of the trace to show that 
the Lagrangian density ${\mathcal L}$ in (\ref{Yang_Mills_action}) is gauge invariant. 

The parameter $g_{\rm YM}$ is the coupling constant.  Dimensional analysis shows that, in four dimensional Yang-Mills theory,  
it is a dimensionless constant and  four-dimensional classical Yang-Mills  theory therefore has no parameters with non-zero scaling dimensions.
It is thus  scale   invariant and it turns out to be conformally invariant.  The scale invariance
generally does not survive quantization, except in some special cases, such as  the maximally supersymmetric Yang-Mills theory which will be of 
central interest to us later on in these lectures.  In that case, the content of the theory in quark fields is carefully tuned to result in maximal
supersymmetry, and at the same time it is scale invariant in both the classical and quantum theories.  We will return to a detailed discussion
of this theory in later lectures. 
 
The equation of motion that results from applying the variational principle to the action of Yang-Mills theory is
\begin{align}\label{ym_fe}
D_\mu F^{\mu\nu}(x)=0
\end{align}
These are the classical Yang-Mills field equations.  Together with the Bianchi identity (\ref{bianchi}), they are the analog of Maxwell's equations
for Yang-Mills theory or, rather, the Yang-Mills theory equations with gauge group $U(1)$ are the source-free Maxwell equations of classical electrodynamics.

 Of most interest to us will be quantized Yang-Mills theory.   We shall take the quantization as being defined by the functional
 integral where, for example, time-ordered correlation functions of gauge invariant operators 
are  obtained as
$$
\left< {\mathcal T}O_1(x_1)\ldots O_n(x_n)\right> = \frac{\int dA_\mu(x) e^{iS[A]} O_1(x_1)\ldots O_n(x_n)}{\int dA_\mu(x) e^{iS[A]}}
$$
where $S$ is the action given in equation (\ref{Yang_Mills_action}) and $O_i(x_i)$ are gauge invariant operators constructed from the connection field,
$A_\mu(x)$ and its deriatives. Here, we have assumed that the
spacetime is Minkowski space. .  If it were Euclidean, rather than Minkowski, 
the functional integral would be
$$
\left< {\mathcal T}O_1(x_1)\ldots O_n(x_n)\right> = \frac{\int dA_\mu(x) e^{-S[A]} O_1(x_1)\ldots O_n(x_n)}{\int dA_\mu(x) e^{-S[A]}}
$$
where ${\mathcal T}$ now denotes Euclidean time ordering. This functional integral is not computable in space-time dimensions greater than two.  Even 
getting an understanding of its salient properties has been a difficult problem which has been intensively researched over the last fifty years and this
research continues today.  The best understood analytic tool is perturbation theory, which is an asymptotic expansion in powers of the coupling $g_{\rm YM}$.
The limit at $g_{\rm YM}=0$ is exactly solvable and perturbation theory gives a systematic technique for computing corrections to that limit when $g_{\rm YM}$
is small enough. However, 
this expansion has many subtleties and, in generic Yang-Mills theory, it can only be applied in a certain kinematic regime.   There is an interesting alternative expansion
of the theory about the limit where $N$ is taken to be large, where $N$ is the rank of the $U(N)$ gauge group.  
However, again, in dimensions greater than two, even the leading term in the large $N$ expansion, that is, the solution of the theory when $N$ is taken to infinity,  is thus far absent.  
What we will learn in the following is that, for certain versions of Yang-Mills theory,
the duality with a certain solutions of string theory will make accessible the leading order of a double expansion, one where a large $N$ limit is taken while
holding the combination $g_{YM}^2N$ fixed, and then a limit of large $g_{YM}^2N$ is taken.  Moreover, the technology for correcting the semi-classical limit 
of string theory should make this limit correctable in a systematic expansion in $1/N$ and in $1/g_{YM}^2N$.  Developing an appreciation of this duality
is the goal of the rest of these lectures.

\section{The Wilson loop}

The matrix $U_C(x_2,x_1)$ is not gauge invariant, it transforms as in equation (\ref{gt4}).  This is so even when $C$ is a closed curve, 
so that the final and initial points are the same ($x_2=x_1$).  
However, in that
case, it transforms by conjugation  $U_C(x,x)\to g(x)U_C(x,x)g^{-1}(x)$ and, even though the matrix itself is not gauge
invariant, we can get a gauge invariant object by taking its trace,  \footnote{In fact, when $U_C(x,x)$ transforms as $U_C(x,x)\to g(x)U_C(x,x)g^{-1}(x)$, its eigenvalues are gauge invariant.  Its
trace is  the sum of its eigenvalues.  We could easily imagine studying more elaborate gauge invariant combinations of  $U_C(x,x)$ 
which give other characteristics of its eigenvalues.}
$${\rm Tr}\left[U_C(x,x)\right] \to {\rm Tr}\left[g(x)U_C(x,x)g^{-1}(x)\right]={\rm Tr}\left[U_C(x,x)\right] $$   

This trace is an important gauge invariant quantity, called the ``Wilson loop'',
\begin{align}\label{wldefinition}
w[C] = {\rm Tr}~U_C(x,x) = {\rm Tr} {\mathcal P} e^{i\oint_C d\tau \dot x^\mu(\tau)A_\mu(x(\tau))}
\end{align}
 The result of the trace, which, although it generally depends on the curve, $C$,  does
not depend on which point on the curve we would choose as the initial and final point.  It is one of the gauge
invariant quantities that it is sometimes useful to  analyze and it, as well as a slight generalization of it, will
play an important role in our study of Yang-Mills theory. 

For example, in the quantized Yang-Mills theory, the Wilson loop has the expectation value
given by the functional integral (in Euclidean space)
\begin{align}\label{wl}
W[C]=\left< w[C] \right> = \frac{\int [dA_\mu(x)] e^{-\int d^4x \frac{1}{2g_{\rm YM}^2}{\rm Tr}F_{\mu\nu}F_{\mu\nu}} 
~ {\rm Tr} {\mathcal P} e^{i\oint_C d\tau \dot x^\mu(\tau)A_\mu(x(\tau))}}
{\int [dA_\mu(x)] e^{-\int d^4x \frac{1}{2g_{\rm YM}^2}{\rm Tr}F_{\mu\nu}F_{\mu\nu}}}
\end{align}

Later on in these lectures, we will be interested in understanding the the dependence of the Wilson loop 
on the coupling constant $g_{\rm YM}$ and the rank of the gauge group, $N$, in a sypersymmetric and conformally
symmetric version of Yang-Mills theory and in some circumstances where these
can be computed.  This will give us a way of studying the duality between the Yang-Mills theory and   string theory. 
Historically, and beyond the scope of what we will discuss in the following, the  behaviour of the Wilson loop for large contours
$C$ has been important in the study of
the dynamical behaviour of  quantized Yang-Mills theory \cite{Wilson:1974sk}. 
For example, a diagnostic of confinement in Yang-Mills theory  is the area law for the Wilson loop.   
For a large contour, $C$,  the area law is the behaviour of the expectation value, 
\begin{align}
W[C]\sim \exp(-\sigma\cdot  {\rm area}[C])
\end{align}
Here, ${\rm area}[C]$ is the
area of a minimal two-dimensional surface where the boundary of that surface is the closed curve, $C$.   The coefficient
of the area, $\sigma$, is a constant with  dimensions of inverse area or energy per unit length. It is called the string tension.

An alternative behaviour, which is characteristic of a deconfined phase of gauge theory,  is the perimeter law, 
\begin{align}
W[C]\sim \exp(- \delta M\cdot {\rm perimeter}[C])
\end{align}
where the perimeter is the length of the curve $C$,  Here, the coefficient, $\delta M$,
 when $\hbar=c=1$, has dimensions of inverse distance, or of energy or mass.  $\delta M$ can be interpreted as contributing to the renormalization
 of the quark mass 
due to interaction of the quark with quantized gauge fields.  
 
In spacetimes with dimensions greater than two, the primary tool for calculating the right-hand-side of 
equation (\ref{wl}) is perturbation theory using Feynman diagrams.  This perturbation theory is super-renormalizable in three dimensions and 
renormalizable in four dimensions.  Three dimensions is an interesting case,  as once a few ultraviolet divergences are dealt with, it is ultraviolet
finite.  No dimensionful scales are generated by renormalization and the only parameter with nonzero scaling dimension is the coupling constant.  
The theory is very strongly coupled in the low energy regime and it is expected to be confining.   
It must therefore have a string tension and dimensional analysis then tells us that the string tension must be proportional to $g_{\rm YM}^4$
 times a dimensionless number. 

In four dimensions, the coupling constant of classical Yang-Mills theory is dimensionless, but it becomes scale-dependent in
the quantized theory.  
 Perturbative calculations can only be reliable in regimes where perturbation theory 
can be trusted, that is, where corrections are controlled
by a small parameter.    In four-dimensional Yang-Mills theory, the coupling 
constant has a non-zero, negative beta-function so that, at 
leading orders it becomes a scale-dependent running coupling, 
\begin{align}\label{runningcouplingofym}
\frac{1}{g_{\rm YM}^2(\mu_1)} -  \frac{1}{g_{\rm YM}^2(\mu_2)} =  {\frac{11}{16\pi^2}\ln (\mu_1^2/\mu_2^2)}
\end{align}
From this formula, we can see that the coupling $g_{\rm YM}(\mu)$ is small and the coupling is weak, so that perturbation theory
is applicable, when the  energy scale $\mu$ is large.   This is the regime where the energy and momentum exchanged in an interaction are large, or,
according to quantum mechanical uncertainty, the time and distance scales involved are small.  The  fact that $g_{\rm YM}(\mu)$  gets
smaller as $\mu$ gets larger is called ``asymptotic freedom''.  

On the other hand, when the scale $\mu$ is decreased, the running coupling gets larger.  This fact, that the 
running coupling is large when the momentum scale, is small is called ``infrared slavery''.   
The latter phenomenon is thought to be at the root of confinement.   However, confining behaviour  is outside of the 
regime where perturbation theory is reliable. The only other computational tool is numerics, to attempt  to perform the integral
in equation (\ref{wl}) numerically.  Considerable progress has been made in this direction, to the point where confinement is a well
established behaviour of three and four spacetime dimensional Yang-Mills theories.   The distance scale where the theory crosses over
from weak coupling to strong coupling  is called the mass scale. This scale, rather than the coupling constant, can be thought of as a
parameter of the quantized Yang-Mills theory.  The other is the $N$ of the $U(N)$ gauge group. 
 
 Quantum chromodynamics (QCD) is the currently widely accepted theory of the strong strong interactions. 
 This model is a Yang-Mills theory with $SU(3)$ gauge group coupled to some fundamental representation massive fermions,
 which are the quarks.     The mass scale of 
QCD is around $250~Mev$. Much of the interesting physics of QCD occurs in the strong coupling regime, at mass scales
below $250~ Mev$, including the formation of the 
hadron spectrum, and practically all of nuclear physics, for example. 
 
The running coupling constant in Yang-Mills theory has implications for the Wilson loop.     Asymptotic freedom implies that Wilson
loops for curves $C$ whose spatial extent and structure are smaller than the inverse mass scale should be computable by perturbation theory.
Those
which are larger are not computable.  As we have already discussed, usually, the larger $C$ regime is the interesting one -- 
for example, the area versus perimeter law
test for confinement is implemented there --
and perturbative Yang-Mills theory cannot be used to compute the Wilson loop in that regime.  
Aside from using computers and numerics to compute the right-hand-side, an endeavour which 
has seen great advances and found a lot of interesting results, there are no reliable analytic techniques.

 It has long been conjectured that the quantized Yang-Mills theory that we have been discussing has a dual description 
 as a string theory \cite{nambu}
 \cite{Makeenko:1979pb}
\cite{Polyakov:1980ca}
\cite{Polyakov:1997tj}.   In that description, the right-hand-side of (\ref{wl}) could be replaced by a string model, where  the
 Wilson loop, for example, would be computed by 
 (in Euclidean space)
\begin{align}\label{wlstring}
  W[C] ~ =~\sum_\chi g_s^{-\chi} \int_{\xi : \delta \xi=C}[d\xi] ~\mu[\xi,\chi] ~e^{-\sigma \cdot{\rm Area}[\xi] }
\end{align}
Here, 
the objects $\xi$, which  are being integrated over,  are two-dimensional surfaces with the property that their boundary
is the curve $C$.    Possible measure factors, which we denote by $\mu[\xi,\chi]$, 
could arise, for example, from integrating out degrees of freedom which live on the surfaces or ghost fields.
 The entity $\chi$  is an integer.   It 
is the topological Euler number ($\chi = 2-2\cdot\#$handles$-\#$boundaries $\leq 2$) of the two-dimensional world-sheet 
of the string,  $\xi$.
  The summand has a Boltzmann factor, $e^{-\sigma \cdot{\rm Area}[\xi] }$ which favours surfaces with minimal area.  
The parameter $\sigma$ has the dimensions of inverse area and it is called the ``string tension''.   In a fundamental string
theory, it is usually denoted as $1/2\pi\alpha'$. 
The dimensionless parameter $g_s$ which controls the   Euler number is called the ``string coupling''.
 If $g_s$ is small, $\xi$ which have
the simplest topologies are emphasized.

The appealing feature of such a model is that it could be computable in an opposite regime to the perturbative regime
where (\ref{wl}) is tractable.  If $g_s$ is small, therefore emphasizing $\xi$'s with disc-like geometry,
and  If $C$ is a large loop, compared with the scale that is set by the inverse of the string tension, then 
the area of a disc which has $C$ as boundary should also be large compared to the inverse string tension, 
  $\sigma $Area$[\xi]>>1$ and the integral would   be computable by semiclassical techniques.   
  The area law and confinement would then be obtained from this model at the semi-classical limit, 
  \begin{align}\label{wlstring1}
  W[C] ~\sim~ \frac{1}{ g_s}~ \mu[\xi_0,\chi=-1] ~e^{-\sigma \cdot {\rm Area}[\xi_0] }
\end{align}
where $\xi_0$ is the minimal surface.
In a confining Yang-Mills theory, one might imagine a duality where
the quantum field theory representation (\ref{wl}) is useful for small Wilson loops, with spatial size smaller than the confinement scale and where
the string model (\ref{wlstring}) is useful for large Wilson loops, with size and features much larger in spatial extent than the confinement scale.    

The existence of such a string representation of the gauge theory is hinted at by the strong coupling expansion of
lattice gauge theory, as well as the behaviour of some lower-dimensional solvable models, like Yang-Mills theory in two dimensions.  Another piece of
evidence in its favour is the large $N$ expansion of Yang-Mills theory, which we will review shortly.   
It organizes the Feynman diagram expansion of the right-hand-side of (\ref{wl}) into a summation over surfaces with given Euler
number and then a summation over Euler numbers, analogous to the right-hand-side of (\ref{wlstring}). 
What one learns from that latter expansion is that 
\begin{align}\label{stringcoupling}
g_s\sim \frac{1}{N}\end{align}
 and that the infinite $N$ limit
projects one to the case where $\xi$ are discs.   In that case, the disc amplitude, 
$$  \int_{\xi : \delta \xi=C}[d\xi]~ \mu[\xi,\chi=-1] ~e^{-\sigma \cdot{\rm Area}[\xi] }$$ 
is identified with
the sum of planar
Feynman diagrams, that is, those diagrams that can be drawn on a plane without crossing lines.  In we will show that this planar limit of Yang-Mills
theory is indeed the result of the appropriate large N limit. 

The subject of the following lectures,  called {\it the AdS/CFT correspondence}, 
contains an explicit example of at least one gauge field theory where
a string model  of the sort outlined in equation (\ref{wlstring}) is indeed realized, 
and where some computations using this duality between gauge fields and strings can be performed \cite{Maldacena:1997re}
\cite{Gubser:1998bc}
\cite{Witten:1998qj}
\cite{Rey:1998ik}
\cite{Maldacena:1998im}. 
Interestingly, the gauge theory in question is not a confining Yang-Mills theory of the type that we have been
discussing.  It is the 
maximally supersymmetric and conformal invariant, ${\mathcal N}=4$ Yang-Mills theory in four-dimensional space-time.  
The Wilson loop is a slight modification of the one which we have described above, adapted to the ${\mathcal N}=4$ theory.   
In this explicit example, the measure
$\mu[\xi,\chi]$ is rather complicated.  However, it has a simple and  remarkable description.  
If we imagine that the discs $\xi$ are not restricted to be embedded in our four-dimensional space-time, but they
are allowed to wander into extra dimensions, so that they live in  the ten-dimensional spacetime, $AdS_5\times S^5$, 
and their boundary is  the curve $C$ which resides in the four-space-time dimensional boundary of $AdS_5$
 the right-hand-side of (\ref{wlstring}) is thought to be an exact representation of (\ref{wl}). 
The space $AdS_5\times S^5$ with a 
constant dilaton and a Ramond-Ramond 4-form field with $N$ units of five-form flux piercing the $S^5$  is a solution of type IIB super-gravity and also IIB string theory.   
$\xi$ are the world-sheets of the critical
type IIB superstring with Dirichlet boundary conditions, so that  $\xi$ has boundary $C$ at the boundary of $AdS_5$.  One could, in principle,  by integrating out
the higher dimensional degrees of freedom, present the theory as four-dimensional, with a measure $\mu[\xi,\chi]$.

This duality, rather than being one between long and short-distance physics, 
is  for a scale invariant theory which does not exhibit quark confinement.  Rather, the coupling constant has vanishing beta function
and it is tuneable.  The duality is then between the weak and strong coupling limits of the theory.  The Yang-Mills quantum field theory is
computable in perturbation theory when the coupling constant is small.  As we shall discuss in the next section, this is so when 
  $g_{\rm YM}^2N<<1$.  The factor of $N$ arises from the multiplicity of particles, or from index sums in Feynman diagrams. 
  The string theory is computable when the string coupling constant is weak, which turns out to be achieved 
  by putting  $g_{\rm YM}\to 0$ and $N\to \infty$ so that $g_{\rm YM}^2N$ is held finite, and when the string tension
  is large, that turns out to be when  $g_{\rm YM}^2N$ is tuned to large values.  

In those limits, the semi-classical limit is an accurate description of the string theory.  Both limits are, 
in principle, solvable in their leading orders and  they are also, in principle, systematically correctable, one
as an expansion in $g_{\rm YM}^2N$ and an asymptotic expansion in $1/N$, the other, also as an asymptotic expansion 
in $1/N$, and then in $1/\sqrt{g_{\rm YM}^2N}$.    
This gives the possibility of using the string theory to study the gauge theory in its strong 
coupling regime, and  using the gauge theory to study a certain strongly coupled 
limit of the string theory.  This feature  is what makes the duality very interesting.   

 The equation (\ref{wlstring}) which describes the string dual of the Wilson loop also has implications for local operators.  In the pure
 Yang-Mills theory that we have been discussion, a Wilson loop for a small circle, of radius $a$, centered on the origin, 
 has an expansion in local operators,
 $$
 W[C] = N + C_4a^4 {\rm Tr} F_{\mu\nu}(0)F^{\mu\nu}(0)+\ldots
 $$
 where $C_4$ is a coefficient.  We could imagine using this expansion to compute a correlation functions of gauge invariant operators by
 considering the expectation values of several small Wilson loops.  Then, the string theory dual calculation would  consider a thin minimal surface which
 connects the locations of the Wilson loops.  This minimal surface is the world-sheet of a  closed string, which, in the low energy limit is
 the propagator of one of the particles in the string spectrum, typically a graviton.  The general duality statement is as follows.  Consider   local operators $O_j(x)$ in ${\mathcal N}=4$ Yang-Mills theory.  We assume that these are operators with good conformal dimensions, that is, they obey commutation relation
 $$
-i \left[D,O_j(0)\right]=\Delta_jO_j(0)
 $$
 with the dilitation generator $D$, 
 where their conformal dimensions are the numbers $\Delta_j$.  (We will supply more details as to the meaning of these in later sections.) 
 These Yang-Mills theory operators are  dual to degrees of freedom, $\phi_j(r,x)$ in the IIB superstring theory on $AdS_5\times S^5$ background. 
 Then, the generating functional for correlations of $O_i(x)$ in the Yang-Mills theory   is expressed in terms
 of the string theory partition function \cite{Gubser:1998bc} \cite{Witten:1998qj}
 $$
 \left<    {\mathcal T} e^{i\int d^4x \varphi_j(x)O_j(x)}\right>
 = Z\left[\lim_{r\to\infty} \phi_j(r,x)=r^{\Delta_j} \varphi_j(x)\right]
 $$
 where the string theory partition function on the right-hand-side is for the string theory with boundary conditions such that each $\phi_j(r,x)$ has the 
 boundary condition that it approaches the classical field $\varphi_i(x)$ as $r$ approaches the boundary of $AdS_5$.

\section{The large N expansion}

The large $N$ expansion is a way or reorganizing the perturbative expansion of a gauge field theory  into
contributions that can be characterized by the topology of the two-dimensional surfaces on which the Feynman diagrams can be
drawn without crossing any lines. In pure Yang-Mills theory, any connected Feynman diagram can be seen to be proportional to 
\begin{align}
\label{simlargeN}
\sim ~  \lambda^{L+\varepsilon/2-1}N^{\chi-\varepsilon/2}
\end{align}
Here, $N$ is the $N$ of the $U(N)$ gauge group of the Yang-Mills theory\footnote{The Yang-Mills theories of interest to us later actually
have an $SU(N)$,  rather than $U(N)$ gauge group.   If it were $SU(N)$, the formula (\ref{simlargeN}) would be modified in a significant
way.  We shall not worry about this difference here, as we will either always consider the infinite $N$ limit, where $U(N)$ and $SU(N)$ are
indeed indistinguishable in the leading order, or the pure Yang-Mills theory where the difference between $U(N)$ and $SU(N)$ is a $U(1)$ 
subgroup whose fields are free fields which decouple from the other Yang-Mills fields and whose contributions in various
cases can be understood and taken into account if it is needed. }   and 
\begin{align}\label{thooft_coupling}
\lambda~\equiv ~g_{\rm YM}^2N
\end{align}
is the combination of the Yang-Mills coupling and $N$ called the 
 't Hooft coupling, The parameter 
$\chi$ is the topological Euler character of the graph (and the Euler character of the 2-dimensional surface on which the graph can be drawn without
crossing any lines).   $L$ is the number of loops in the graph and $\varepsilon$ is the number of external lines. 
  We shall derive the  expression (\ref{simlargeN}) shortly.

\begin{figure}
 ~~~~~~~~~~~~~~~~~~~~~~~~~~~~~\includegraphics[scale=.5]{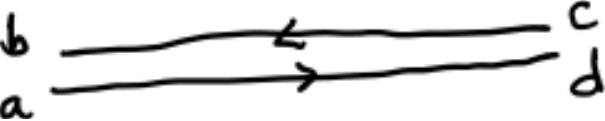}\\
\begin{caption} {  The propagator of the Yang-Mills field in fat graph notation has two lines with
indices which are connected by the lines identitied as depicted.    \label{rules1}  
}\end{caption}
 \end{figure}  
 
 \begin{figure}
 ~~~~~~~~~~~~~~~~~~~~~~~~~~~~~\includegraphics[scale=.5]{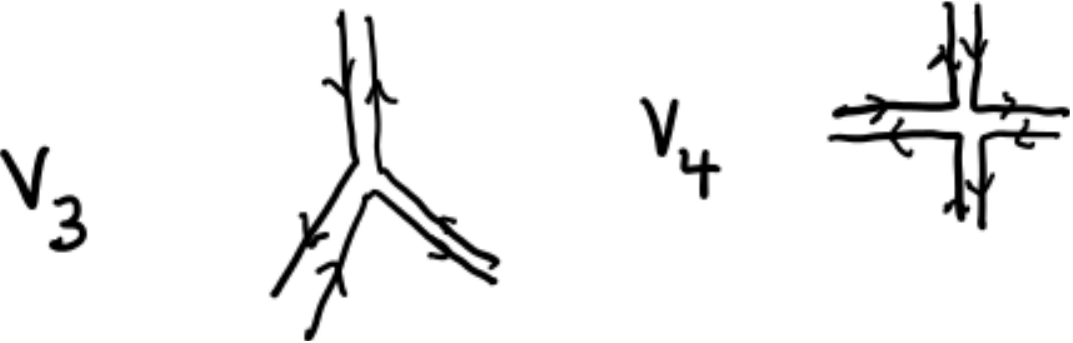}\\
\begin{caption} {   The three and four-point vertices of Yang-Mills theory in fat graph notation.   \label{rules4}  
}\end{caption}
 \end{figure}

When we take the large $N$ limit,  where we put $N\to\infty$ while holding the 't Hooft coupling, $\lambda$, and the number of loops $L$ and the number
of external lines $\varepsilon$ constant,  equation (\ref{simlargeN}) tells us that the dominant order is the sum of all Feynman diagrams with maximum Euler character, $\chi$.  

If we consider vacuum diagrams, with no external lines, so that $\varepsilon=0$, the maximum Euler character is $\chi=2$, the Euler number of the 2-sphere and the Feynman diagrams of this order can be drawn on a 2-sphere without crossing lines.     

If we consider  a contribution with external lines, $\varepsilon>0$,  the faces of the diagram into which the the external lines are  emitted are  
 counted as holes, rather than faces.   The Euler character is maximal if the number of holes is minimal.  The surface can generally have
 only one hole. (
 To be clear, in this
 case, all emissions of external lines are
into a single  hole.   )
The  2-sphere with a hole cut out of it has $\chi=1$ and these diagrams are of order $N^{1-\varepsilon/2}$.
 The Riemann surface on which the graphs of this leading order can be drawn has $\chi=1$ and the topology of the plane. The
graphs are therefore called planar and the truncation of the full Yang-Mills theory to the sum over planar graphs for any multi-point function
is called the planar limit of Yang-Mills theory. 

This leading order, planar, contribution  
contains Feynman diagrams to all orders in the usual perturbative loop expansion in the Yang-Mills coupling constant. The planar limit of Yang-Mills theory  therefore 
  appears to still be a highly non-trivial dynamical system in its
own right. The diagrams
are weighted by the 't Hooft coupling $\lambda$ to the power of the number of loops.  Thus, the weak coupling limit of this planar theory ccurs where $\lambda$ 
is small, in that small $\lambda$ favours
diagrams with smaller numbers of loops whereas strong coupling is large $\lambda$ in that it favours diagrams with large numbers of loops. 
  Before AdS/CFT, even this leading, planar limit of Yang-Mills theory was not solvable for Yang-Mills theory in spacetime dimensions greater than two.
 With the advent of AdS/CFT, the strong coupling limit of the planar limit of the four dimensional ${\mathcal N}=4$ supersymmetric Yang-Mills 
 theory and some field theories which are related to it  
 have become solvable in many circumstances.  In the following, we shall define what we mean by ``strong coupling limit''
 in this context.

The analog of the large N expansion in statistical models  with matrix-valued degrees of freedom had been studied for a long time, particularly 
in combinatorics \cite{tutte1} \cite{tutte2} and the statistical mechanical theories of triangulated random surfaces \cite{Kazakov:1985ea}. The application of the idea to Yang-Mills theory is due to t'Hooft and dates back to the 1970's  \cite{tHooft:1973alw}  \cite{Brezin:1977sv}.  The motivation was to understand in what way Yang-Mills theory could look like a string theory.   The large $N$ expansion is indeed one way: reorganize the perturbation theory into one that resembles the string perturbative expansion of quantities in the closed string sigma model in powers of the string coupling constant $g_s$.  In the string sigma model,  the expansion is in the powers  $g_s^{-\chi}$ where $\chi$ is  the Euler character of the world-sheets of the strings. There, the leading order contains strings with world-sheets which are topologically equivalent to the two-sphere.   

We now want to find a basic derivation of some of the above statements about the large $N$ limit.  
 In order to discuss perturbation theory, we  should consider the gauge-fixed Lagrangian density 
\begin{align}
{\mathcal L}_{\rm gf} =  {\rm Tr}\left\{ - \partial_\mu A_\nu\partial^\mu A^\nu + \left(1- \frac{1}{\xi}\right)(\partial_\mu A^\mu)^2+
\partial_\mu \bar c\partial^\mu c - ig_{\rm YM} \partial_\mu\bar c \left[ A^\mu,c\right]-   \right.   \nonumber    \\
\left. - {i}g_{\rm YM}  \partial_\mu A_\nu[A^\mu,A^\nu]
 +\frac{1}{2 }g^2_{\rm YM} [A_\mu,A_\nu][A^\mu,A^\nu]
\right\}
\label{lgf}\end{align}
 where $\xi$ is the gauge fixing parameter and  $c$ and $\bar c$ are Faddeev-Popov ghosts, which are also matrix-valued fields. 
 Here, we have scaled the fields so that the Yang-Mills coupling constant appears in front of the Lagrangian density. 
  
 To get the large $N$ expansion, it is convenient to adopt a fat graph notation for the propagators of matrix-valued fields.
 The fat graph  for the propagator is depicted in figure \ref{rules1} and for the three- and four-point vertices is depicted in figure \ref{rules4}.
The propagators of matrix-valued fields carry the matrix indices.  For example, the gluon propagator in the Feynman gauge ($\xi=1$)
 and the Fadeev-Popov ghost propagator are
\begin{align}
\left< A^\mu_{ab}(x)A^\nu_{cd}(y)\right>_{g_{\rm YM}=0}=~ {\delta_{ad}}\delta_{bc}
~\int \frac{d^4k}{(2\pi)^4}e^{ik_\mu(x-y)^\mu} \frac{-i\eta^{\mu\nu} }{k^2-i\epsilon}
\label{gluon_propagator}\\
\left< c_{ab}(x)\bar c_{cd}(y)\right>_{g_{\rm YM}=0}=~ {\delta_{ad}}\delta_{bc}
~\int \frac{d^4k}{(2\pi)^4}e^{ik_\mu(x-y)^\mu} \frac{i }{k^2-i\epsilon}
\label{ghost_propagator}
\end{align}
The feature of the propagators that is important to us is  the delta-functions with matrix indices on the right-hand-side.  
They are a result of global U(N) symmetry that is left over after gauge fixing and they are appropriate to $U(N)$ gauge theory.
For SU(N), $ {\delta_{ad}}\delta_{bc}$ would be replaced by $ {\delta_{ad}}\delta_{bc} -\frac{1}{N}{\delta_{ab}}\delta_{cd}$ and the large $N$ expansion would be more complicated, although the differences would not be seen in the leading order.    In the following, for simplicity, we shall mostly be interested in $U(N)$. 
The solid lines in the propagator in figure \ref{rules1} connect the indices that are identified by the delta functions in equations (\ref{gluon_propagator}) and
(\ref{ghost_propagator}). 
 
 Vertices are gotten from  using these propagators  in Wick contractions with the interaction terms in  ($i$ times )  the action
\begin{align}
&{g_{\rm YM}}{\rm Tr} \partial_\mu\bar c \left[ A^\mu,c\right]   \label{vertex1}    \\
&{g_{\rm YM}}{\rm Tr}   \partial_\mu A_\nu[A^\mu,A^\nu]    \label{vertex2}   \\
&{g_{\rm YM}^2}{\rm Tr}\frac{i}{2 } [A_\mu,A_\nu][A^\mu,A^\nu]  \label{vertex3}
\end{align}
The fat graph notation for the vertices is depicted in figure \ref{rules4}.

We are interested in studying the dependence of generic Feynman diagrams on $N$.  
The counting of powers  of N in the large N expansion is best seen by using the fat graph 
notation that we have outlined above.  
  


\begin{figure}
 ~~~~~~~~~~~~~~~~~~~~~~~~~~~~~\includegraphics[scale=.6]{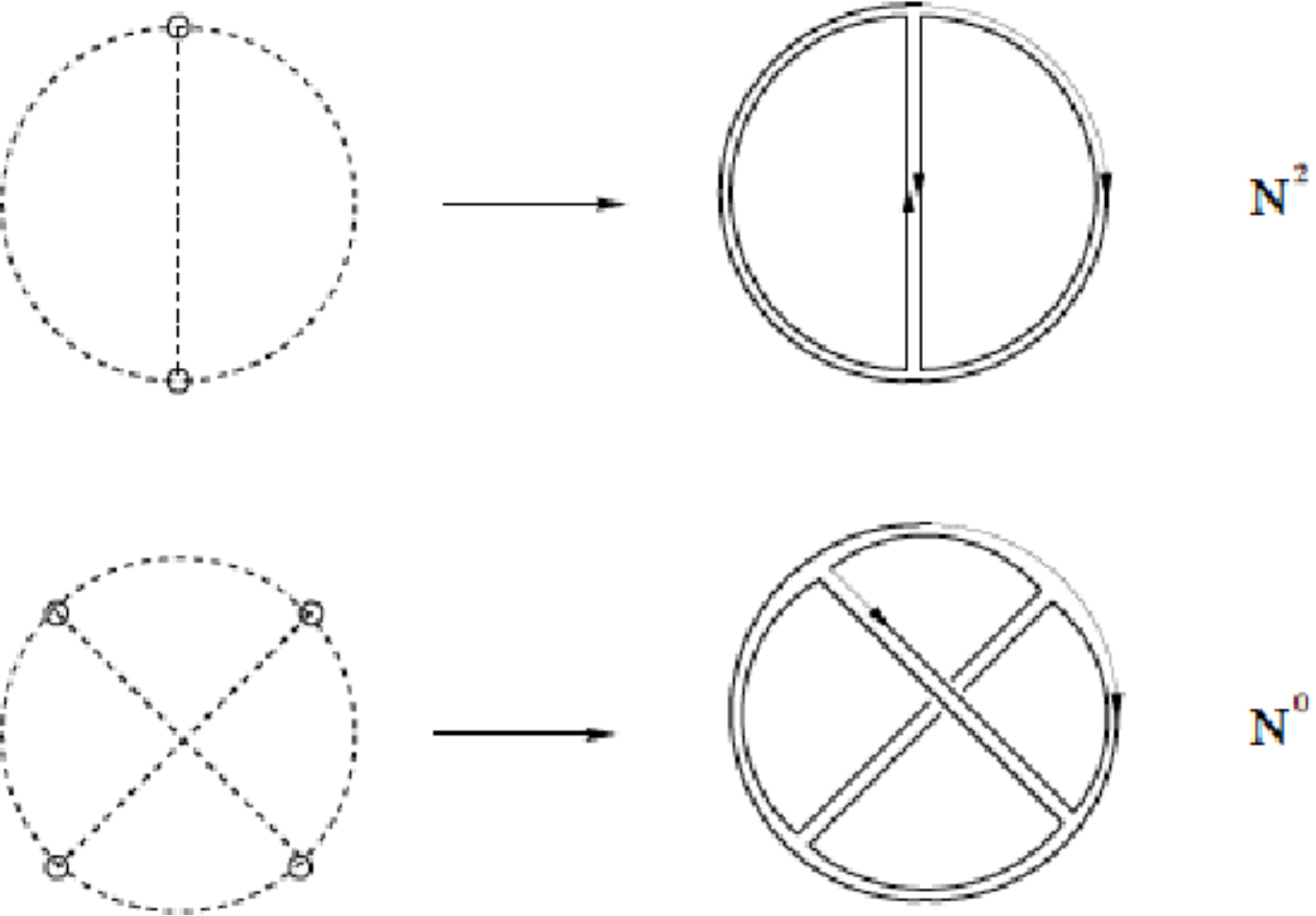}\\
\begin{caption} {Some examples of Feynman diagrams which would be used to calculate the vacuum amplitude.  The graphs on the
right are equvalent to the graphs on the left, but they are written in fat-graph notation.   Once the graph is depicted that way, the number of
closed index lines is equal to the number of faces of the graph.  The sum over indices in each closed index line produces a factor of $N$. The top
graph has two loops and it is ``planar'' in that it has been drawn on the plane without crossing any lines.  
It has three faces which contribute a factor of $N^3$.  (The exterior of the graph counts as a face.)  
It also has two vertices, which together give it a factor of $g_{\rm YM}^2$. The total factor with the graph is therefore $g_{\rm YM}^2N^3=\lambda N^2$.
Since the graph is planar, and it has no external lines, its Euler character is that of the 2-sphere, $\chi=2$. 
The second graph has four loops and it is not planar.  It has two index loops, so it has the factor  $g_{\rm YM}^4N^2=\lambda^2 $.  Since it can be drawn on a torus without crossing lines, its Euler character is $\chi=0$.   The factors of both
graphs agree with the general formula in equation (\ref{simlargeN}), $ \lambda^{L+\varepsilon/2-1}N^{\chi-\varepsilon/2}$,  when we put the number of external lines $\varepsilon$ to zero.
     \label{ex}
}\end{caption}
 \end{figure}

\begin{figure}
 ~~~~~~~~~~~~~~~~~~~~~~~~~~~~~\includegraphics[scale=.5]{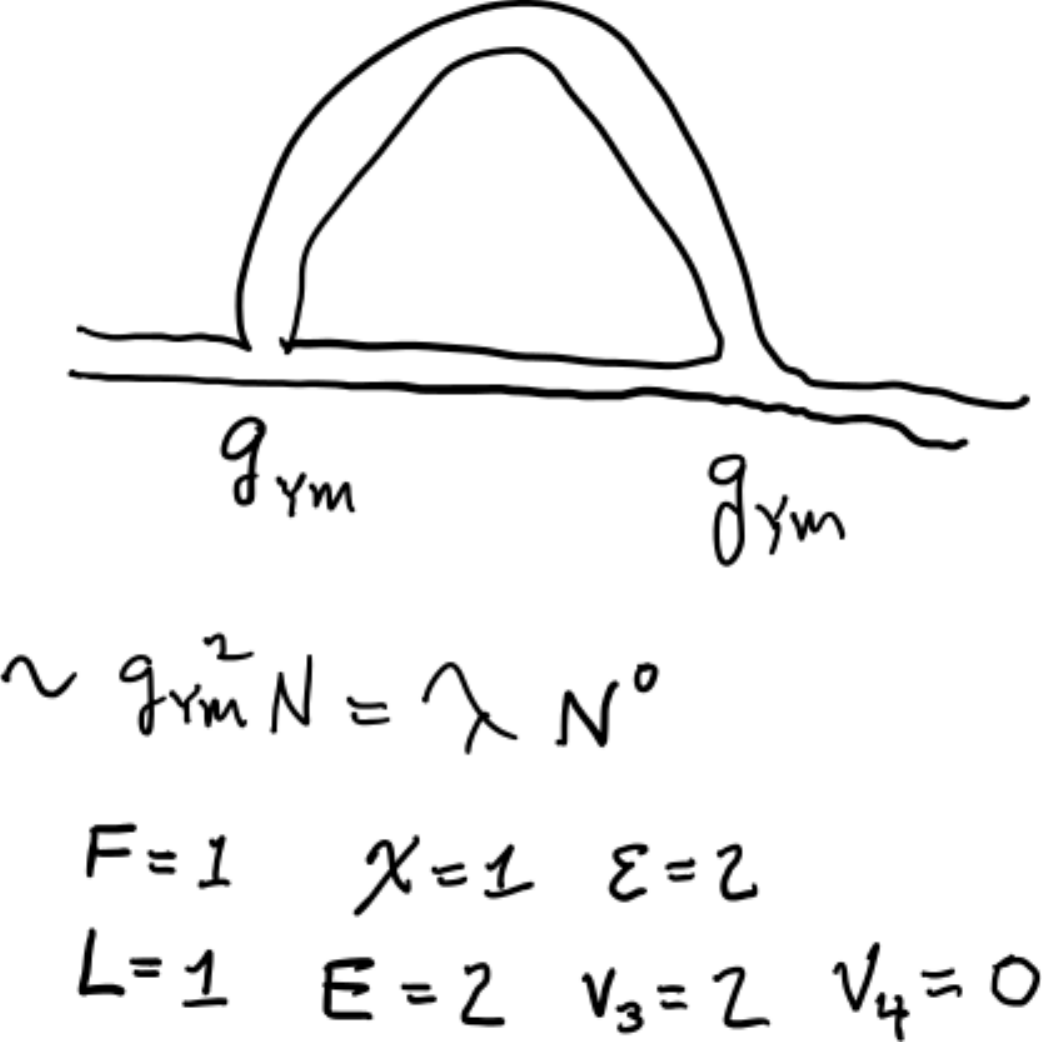}\\
\begin{caption} {  A one-loop correction to the Yang-Mills field propagator is depicted. 
The graph has two internal lines, thus two edges, $E=2$.  It has
two three-point vertices and no four-point vertices, thus, $V_3=2$, $V_4=0$.  It has one face and thus $F=1$.  The index sum corresponding to
the face produces a factor of $N$.  
 The Euler character of this graph is that of the plane, $\chi=1$.  To see this, one
can consider the entire exterior of the graph as a hole into which the two external lines are emitted.Tthe number of loops is $\ell=1$.  The counting displayed in the figure
yields $\lambda N^0$ which agrees with equation (\ref{simlargeN}), $ \lambda^{L+\varepsilon/2-1}N^{\chi-\varepsilon/2}$.    \label{rules2}  
}\end{caption}
 \end{figure}  
  
\begin{figure}
 ~~~~~~~~~~~~~~~~~~~~~~~~~~~~~\includegraphics[scale=.5]{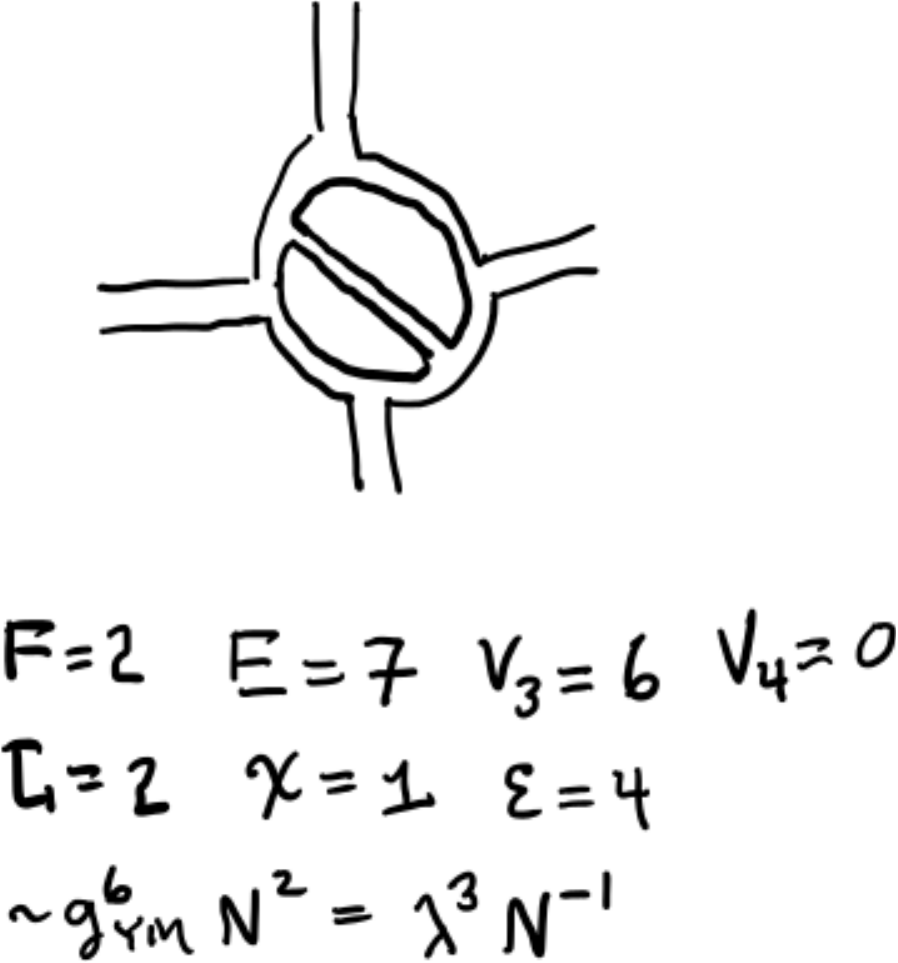}\\
\begin{caption} {    A two-loop correction to the Yang-Mills field four-point function is depicted. The graph has seven internal lines, thus seven edges, $E=7$.  It has
six three-point vertices and no four-point vertices, thus, $V_3=6$, $V_4=0$.  It has two faces and thus $F=2$.  The two index sums produce a factor of $N^2$.  
 The Euler character of this graph is that of the plane, $\chi=1$. To see this, one
can consider the entire exterior of the graph as a hole into which the four external lines are emitted. The number of external lines is $\varepsilon=4$ and the number of loops is $\ell=2$.  
The counting displayed in the figure
yields a factor of $\lambda/N$ which agrees with what we would obtain from equation (\ref{simlargeN}), $ \lambda^{L+\varepsilon/2-1}N^{\chi-\varepsilon/2}$.     \label{rules3}  
}\end{caption}
 \end{figure}

 Consider a connected Feynman diagram which contributes to an $\varepsilon$-point correlation function.  
 We shall call it a ``graph''.
 It is an assembly of vertices and of propagators connecting
 vertices to each other as well as to the external points. 

 Let us say that there are $V_3$ three-point vertices and $V_4$ four-point vertices. 
  The internal double  lines of the graph form `edges''.  Let us say that there are $E$ edges.  As well, there must be $\varepsilon$ external lines, that is, lines which connect each of the $\varepsilon$ external points to vertices which are in the graph. 

Now, let us find the factors of $g_{\rm YM}$ and of $N$ which must go with a given graph.   
 \begin{enumerate}
\item{}Each three-point vertex of a graph is a vertex from which  three  lines emanate.  Each three-point 
vertex is accompanied by a factor of ${g_{\rm YM}}$ and the three-point vertices therefore contribute the factor
$$
 ( g_{\rm YM})^{V_3}
$$
\item{}Each four-point vertex  is a point from which  four lines emanate.  Each   is accompanied by a factor of ${g_{\rm YM}}^2$.
The four-point vertices  contribute the factor
$$
( g_{\rm YM})^{2V_4}
$$
\item{}Each face yields a factor of $N$ coming from the summation over indices in a closed line,
resulting in the factor
$$
(N)^F
$$
\item{}
A summary of the factors so far is the product of the above three quantities, 
$$
  (g_{\rm YM})^{V_3+2V_4}(N)^F
  $$
\end{enumerate}
Now, there are some identities that the graph must obey:
\begin{enumerate}
\item{}Each three point vertex is a source of three lines and each four-point vertex is a source of four lines.  Each internal line must
attach to two vertices, each external line to one vertex.  This relates the numbers of vertices and lines as
$$
\varepsilon+2E=3V_3+4V_4
$$
\item{}The number of loops is equal to the number of momentum space volume integrals which need to be done
in order to evaluate the Feynman graph.  Each external line has a predetermined momentum, so they are not counted. 
Each internal line has a momentum and each vertex has a momentum conserving delta function.  When all momentum 
conserving delta functions are taken into account, one must be left over to conserve momentum in the entire graph.
Thus, the net number of momentum space volume integrals is 
$$
L=E -V_3-V_4+1
$$
\item{}The Euler character $\chi$ is a topological characteristic of the graph.   
It is equal to the number of faces minus the number of edges plus the total number
of vertices
$$
\chi=F-E+(V_3+V_4)
$$
This is also a topological characteristic of the   two-dimensional surface
on which the graph can be drawn with out crossing any lines.   
For the  surface, this is
$$
  \chi = 2-2g-h
$$
  where $g$ is the  number of handles on the surface, and $h$ is the number of holes in the surface.   Both of these are non-negative integers, so
  the maximum Euler number is two, that of the sphere, $\chi_{\rm sphere}=2$.  A torus
  has one handle and no holes and it has Euler character $\chi_{\rm torus}=0$.  A disc has no handles and one hole and
  it has $\chi_{\rm disc}=1$.  An annulus has $\chi_{\rm annulus}=0$. In summary,
  $$
  \begin{matrix} ~~ &  {\rm sphere}& {\rm disc} &{\rm torus} & {\rm cylinder}& \ldots \cr 
  -\chi & -2 & -1 & 0 & 0 & \ldots \cr \end{matrix}
  $$
\end{enumerate}
Putting these together, some simple algebra tells us that a Feynman graph is weighted by
the overall factor
\begin{align}\label{box}
\boxed{ \lambda^{L+\varepsilon/2-1}N^{\chi-\varepsilon/2}\cdot{\rm Feynman~integral}~~}
 \end{align}
 which is the expression that we used in equation 
(\ref{simlargeN}). 
 
 \subsection{Large $N$ factorization}    
  
There is one rather remarkable result of the large $N$ expansion, which is called large $N$ factorization.  If, in the Yang-Mills theory, 
we compute the correlation function of any number of $U(N)$  invariant operators, 
\begin{align}
 \left< {\mathcal O}_1(x_1) {\mathcal O}_2(x_2)\ldots {\mathcal O}_k(x_k)\right>
=\left< {\mathcal O}_1(x_1)\right> \left< {\mathcal O}_2(x_2)  \right> \ldots\left<{\mathcal O}_k(x_k)\right> 
\left\{1+{\mathcal O}\left(\frac{1}{N^2}\right)\right\}
\end{align}
This property is called large $N$ factorization.  In the leading order, $SU(N)$ invariant operators behave as if they
are classical fields,  they are uncorrelated. Of course, in the Ads/CFT correspondence which we shall describe shortly, the classical variables
are classical string theory degrees of freedom.  The large $N$ limit is the one which turns off the string interactions, that is, amplitudes where strings split
or join.  However, even
classical, or more accurately, tree-level string theory is technically very complicated and we shall need another limit to make it solvable.  We will
come back to this subject in later sections.

It is easy to prove factorization if we make some assumptions about the operators.   Let us assume that they are 
gauge invariant products of the matrix-valued Yang-Mills fields and the derivatives of Yang-Mills fields, all evaluated at the same point, and that they are formed by 
taking a single trace a product of such matrices.  An example would be ${\rm Tr}F_{\mu\nu}(x)F^{\mu\nu}(x)$   or   ${\rm Tr}F_{\mu\nu}(x)F^{\nu\rho}(x)F_\rho^{~~\mu}(x)$.  We could then add them to the Lagrangian density of Yang-Mills theory with position-dependent coefficients, 
$$
{\mathcal L}\to {\mathcal L}+\sum_i g_i(x){\mathcal O}_i(x)
$$
and treat them as vertices.   Then, a generalization of the arguments above which includes the new vertices (and did not depend on the coupling
constants actually being constants) would tell us
that the dominant diagrams contributing to $\left< {\mathcal O}_i(x_i)\right>$ are  the connected planar diagrams, which are of order $N^2$.  These are just the
leading part of the connected vacuum
amplitude of the Yang-Mills theory with the generalized vertices included, and then Taylor expanded to leading oder in $g_i(x_i)$,  and with the other $g_i$'s set to zero.
Thus, the leading order connected diagrams in $\left< {\mathcal O}_i(x_i)\right>$ are of order $N^2$.   \footnote{Of course, non-connected diagrams cancel with the
expansion of the denominator in the expression
$$
\left< {\mathcal O}_i(x_i)\right> = \frac{ \int [dA] e^{-S}  {\mathcal O}_i(x_i)}{ \int [dA] e^{-S} }
$$
The digrams which do not cancel are those which are connected to the point $x_i$. This is one of Goldstone's theorems.}

Moreover, the leading order of $\left< {\mathcal O}_1(x_1)\right> \left< {\mathcal O}_2(x_2)  \right>$ would be $N^4$.    On the other hand, 
the connected two-point function 
$$
 \left< {\mathcal O}_1(x_1)  {\mathcal O}_2(x_2)  \right>_C = 
  \left< {\mathcal O}_1(x_1)  {\mathcal O}_2(x_2)  \right> - 
   \left< {\mathcal O}_1(x_1)\right> \left< {\mathcal O}_2(x_2)  \right>
   $$
   would be the coefficient of $ g_1(x_1) g_2(x_2)$ in the functional Taylor expansion  in the $g_i$'s  of .the sum of connected planar vacuum diagrams 
   for the Yang-Mills theory with enhanced Lagrangian density, $ {\mathcal L}+\sum_i g_i(x){\mathcal O}_i(x)$.  As we have already argued, the leading order
   contributions to this quantity must be of order $N^2$.   Thus, in summary, 
$$
   \left< {\mathcal O}_1(x_1)\right> \left< {\mathcal O}_2(x_2)  \right>\sim N^4
   $$
   and
  $$
  \left< {\mathcal O}_1(x_1)  {\mathcal O}_2(x_2)  \right> - 
   \left< {\mathcal O}_1(x_1)\right> \left< {\mathcal O}_2(x_2)  \right>\sim N^2
   $$ 
 and
 $$
  \left< {\mathcal O}_1(x_1)  {\mathcal O}_2(x_2)  \right>= 
   \left< {\mathcal O}_1(x_1)\right> \left< {\mathcal O}_2(x_2)  \right>\left\{1+{\mathcal O}(1/N^2) \right\}
   $$
   It is easy to generalize this argument to higher order correlation functions.  The upshot is that. at large $N$, 
   single-trace gauge invariant local operators are uncorrelated.   It is also easy to generalize to other operators,
   like correlators of Wilson loops. 
     
There are several other important facts that are related to factorization  and that I will leave it to the reader to derive, or find in the literature.
An important one is what happens when we couple the Yang-Mills theory to  a quark which transforms in the fundamental representation of
the gauge group,  that is,  a quark field which, instead of being matrix-valued like the Yang-Mills field, is a complex vector, with only one rather than two
of the gauge group indices.   In the fat-graph notation, the propagator of such a field contains only a single line.  When they occur in Feynman
diagrams, these propagators must form boundaries of the graph.   For example, a closed internal loop with a single line propagator should be
interpreted as a hole in the graph.  It therefore has smaller Euler character than the graphs without the internal loop and it is therefore subdominant
in the large $N$ expansion. For this reason, fundamental representation quarks do not contribute to the large $N$ limit of the vacuum energy of a 
Yang-Mills theory which is coupled to them.   

\section{D branes, black branes and the Maldacena conjecture}

\subsection{D3 branes and black D3 branes}

The AdS/CFT correspondence is a duality between string theory on asymptotically anti-de Sitter background space-times
and conformal field theories defined on flat space of dimensions  
one smaller than the anti-de Sitter space. These theories are
``holographic'' in the sense that  all of the information that is needed to specify a state of a string theory, which contains a theory of quantum gravity, 
 is encoded in the quantum state of a  quantum field theory on flat space-time and in a lower dimension.  
   The quantum field theory state is the hologram.    This is a generalization of ideas which stemmed from 
the use of black hole entropy to count the degrees  of freedom in quantum gravity \cite{tHooft:1993dmi}
 \cite{Susskind:1994vu} \cite{Bousso:2002ju}.

This duality can hold at various levels, and there are several examples.   Here, we will consider the duality 
in its best established example,  
the mapping between ${\mathcal N}=4$ supersymmetric Yang-Mills 
theory\footnote{We will call it ${\mathcal N}=4$
theory for short.} in  four flat space-time dimensions and 
type IIB superstring theory on the ten dimensional $AdS_5\times S^5$ background.  The ${\mathcal N}=4$ theory is a superconformal field theory, thus 
the CFT in AdS/CFT.  We will also consider the duality in its strongest form, where
it is a one-to-one mapping of all of the quantum states, operators, observables, processes, et cetera, between the two theories.   For example, the 
vacuum state of the ${\mathcal N}=4$ theory corresponds to the state of the string theory which is the empty $AdS_5\otimes S^5$ geometry, without
the excitation of gravitons or other modes in the string theory spectrum.  

The ${\mathcal N}=4$ theory has two parameters, the dimensionless
coupling constant $g_{\rm YM}$ which governs the strength of interactions,
 and the $N$ of the $SU(N)$ gauge group.   These are combined to form the 't Hooft coupling
\begin{align}\label{thooft_coupling}
\lambda \equiv  g_{\rm YM}^2 N
\end{align}
 which plays an important role in the large $N$ limit.  
 
 The string theory also has two parameters.
One of them is the string coupling constant, $g_s$ which regulates the strength of string interactions, that is, the propensity of 
strings to split and join.
The other parameter is the string tension, which is effectively the radius of curvature of the geometry in which the string is embedded, written
in units of fundamental string tension $2\pi\alpha'$.  The $AdS_5$ and $S^5$ are constant negative and positive curvature spaces, respectively,
which have the same radii of curvature which we call $L$. 
  Fundamental strings also have a string tension, $\alpha'$ which has the units of length squared and the dimensionless
parameter is $L/\sqrt{\alpha'}$.  

The parameters of ${\mathcal N}=4$ theory and the IIB string are mapped onto each other as
\begin{align}
4\pi g_s &= g_{\rm YM}^2 \label{couplings_duality}\\
\frac{L^4}{{\alpha'}^2} &=  \lambda    \equiv  g_{\rm YM}^2N   \label{curvature_lambda}
\end{align}
With these two identifications, we are supposed to be able to find a one-to-one mapping of all of the attributes of the gauge field theory to the string
theory and vice-versa.   Of course, the precise nature of this mapping in all of its detail is still a subject that is undergoing development.  
 One of the fascinating frontiers of this subject is the project of filling in the details of exactly what is dual to what. 
 
 What makes this mapping interesting is the fact that it is a weak coupling to strong coupling duality.   The  gauge field theory is tractable in the limit
 where both $g_{\rm YM}$ and $\lambda$ are small, so that perturbation theory, which is practically the only analytic approach, is accurate.  
 The string theory, on the other hand, is tractable where it is semi-classical, 
 in the limit where $g_s$ is small, and the limit of weak curvature, $\frac{L}{\sqrt{\alpha'}}\to\infty$.

 Of course, the duality in its strongest form, and even in the weaker forms where it is only approximate, is only conjectured to exist.  
 There is no proof of it at the mathematical level.  It is 
 not even clear that there can be such a proof, as it would require a mathematically precise definition of the theories on both sides
 of the mapping. A weaker version of the duality would have it holding only in the large $N$ limit or only in the weak curvature $\frac{L}{\sqrt{\alpha'}}\to\infty$
 limit.  Both of these are still very interesting as, even an  understanding of non-trivial gauge field theories in the large $N$ limit, and tree level
 superstring theory is for from complete thus far.   The weak curvature limit would still allow us to use the duality to compute the strong coupiling
 limit of the quantum field theory. 

\subsection{Weakly coupled string theory with D3 branes}

Let us first begin with weakly coupled IIB string theory on ten-dimensional Minkowski space
with $N$ flat, infinite D3 branes occupying the cartesian space-time dimensions $x^0,x^1,x^2,x^3$ 
as 
$$
\begin{array}{ccccccccccc} ~{\rm dimension}~ & 0 & 1 & 2 & 3 & 4 & 5 & 6 & 7 & 8 & 9 \cr  ~~~~\cr
                N~D3 & X & X & X & X & O & O & O & O & O & O \cr
               \end{array}
$$
In the table above, the D3 branes fill the dimensions marked by $X$ and they 
sit at points in the dimensions marked $0$. They are distributed this way, parallel, flat and infinitely extended, because this configuration should be a local minimum of the energy.  Also, when they have coinciding positions, they violate as few symmetries as possible. 
They are $\frac{1}{2}$-BPS objects which  reduce the number of supersymmetries of the IIB string theory from 32 to 16. 
They also have Poincare symmetry in the dimensions that they occupy, $x^0,x^1,x^2,x^3$. 
The transverse directions $x^4,...,x^9$ have an $SO(6)$ (=$SU(4)$) symmetry
under rotating these coordinates into each other. These symmetries will be enhanced to superconformal symmetry in
the low energy limit.  This leads to a doubling of the supersymmetries, back to 32, which is needed for the equivalence
to the full IIB string theory on $AdS_5\times S^5$. 

This system contains closed strings moving in ten dimensional space-time 
surrounding the D3 branes and open strings which begin and end on the D3 branes. 
The string coupling constant $g_s$ governs the propensity of strings to interact by splitting
and joining.  This interaction will be suppressed if this coupling constant is small,  $g_s<<1$.  
However, this is not quite enough to guarantee that the interactions are weak.  The open strings can begin or end on any one of the $N$  D3 branes.
Summing over these possibilities can produce factors of $N$.   
These factors will occur in such a way that the effective coupling constant is $g_sN$.  Thus, the open string
sector of this system will be weakly coupled and perturbative string theory will give an accurate
description of this system only when $g_sN<<1$. 
Of course, this also guarantees that $g_s<<1$. 

Thus, when $g_sN<<1$, the system  is accurately described by these weakly interacting
strings on flat ten-dimensional spacetime. 
In this weak coupling limit, both the open and closed strings
have states which behave as massless relativistic particles, as well as infinite towers of 
massive relativistic particles with masses $M^2\sim\frac{1}{\alpha'}$.

Now, let us take the low energy limit. 
 In this limit, the massive states of both the open and closed strings should decouple.  Intuitively, ``taking the low energy limit''
 is simply agreeing that we restrict our ability to
probe the system so that we are only allowed low frequency, large wavelength probes,
to the extent that we can never disturb it violently enough to excite a massive state of a string.  
The low energy limit of the closed string sector of IIB string theory is IIB super-gravity and, in this background of 
flat ten-dimensional Minkowski space, the low energy excitations of super-gravity are the fields of the graviton super-multiplet which
contains the graviton, dilaton, anti-symmetric tensor field and their superpartners.  
Thus, the closed string degrees of freedom appear    
as weakly interacting super-gravitons propagating in the bulk of the flat ten dimensional spacetime surrounding the D3 branes.  
Their interactions with each other and with the open strings are  weak since the strength of the interactions  scale like powers of the ten-dimensional 
Planck length times the energies of the strings,  
$$g_s^2(\alpha')^4\cdot{\rm (energy)}^8$$ which is small because $g_s$ is small and the energy is much smaller than $1/\sqrt{\alpha'}$. 
 

As well as the closed strings, there are open strings which begin and end on the D3 branes.   The massless modes of these open strings are the quantum fields of the ${\mathcal N}=4$ theory, living on the world-volume of the D3 brane.  In the low energy limit, they  decouple
from the massive degrees of freedom of the open and closed strings.  However, they do not become free field theory in the low energy
limit.  The ${\mathcal N}=4$ theory is conformally invariant and scaling to low energies does not change the strength of the interactions.  It remains
a non-trivial interacting quantum field theory in the low energy limit.  The coupling constant of the gauge field theory is directly related to the 
coupling constant of the string theory, 
\begin{align}
g_{\rm YM}^2~=~4\pi g_s
\label{g_YM}
\end{align}

Our conclusion is that, in the limits $g_sN<<1$ and where we keep states of the theory where energies are much smaller than $\frac{1}{\sqrt{\alpha'}}$,
the remaining dynamical system consists of free super-gravitons propagating in flat ten dimensional spacetime and fully interacting four space-time dimensional
${\mathcal N}=4$ theory. What is more, the ${\mathcal N}=4$ theory is weakly coupled and one should be able to do accurate calculations of quantities like correlation functions using perturbation theory, that is, in an asymptotic expansion in $g_{\rm YM}^2N$ about free field theory.


 \subsection{The black D3 brane solution of IIB super-gravity}
 
 Now, let us turn to another description of the stack of $N$ D3 branes, that of a Ramond-Ramond-charged state of IIB super-gravity
 which has, as a classical solution,  a black D3 brane geometry with N units of Ramond-Ramond four-form charge.  
 It is actually a conjecture that the D3 branes can be described
 in this way.  The description is accurate  in a regime that is different from the weak coupling one which we have discussed above. 
 Since it is in a different regime, and there is no accurate description of the system in the intermediate range, it is not possible to follow the
 system adiabatically from one limit to the other and therefore no rigorous proof that they describe the same system.  
 On the other hand, the equivalence,
 which was first realized by Polchinski \cite{Polchinski:1995mt},
 fits so well with so much of what is known of the states of string theory, there is little doubt that it
 is correct.  
 
 The low energy effective action for the type IIB superstring is the action of type IIB super-gravity, whose bosonic part is
\begin{align}\label{sugra}
S_{\rm SUGRA} = \frac{1}{(2\pi)^7{\alpha'}^4}\int d^{10}x \sqrt{-g}\left\{ e^{-2\varphi} \left(R +4(\nabla\varphi)^2\right)
-\sum_p\frac{2}{(8-p)!}F_{p+2}^2\right\}
\end{align}
This theory contains the metric tensor, $g_{AB}(x)$, the dilaton,  $\varphi(x)$,  and p+2-form fields $F_{p+2}(x)=dA_{p+1}(x)$,  
the field strengths for the Ramond-Ramond (p+1)-form fields $A_{p+1}(x) $.  Type IIB super-gravity which we are interested
in here has only p-odd forms whereas
the IIA super-gravity would have only p-even forms. Moreover, in the IIB theory, the five-form $F_5$ is constrained
to be self-dual, $F_5=*F_5$ where the star denotes the Hodge dual. This constraint is usually imposed at the level of
the equations of motion which are derived from (\ref{sugra}) using a variational principle.  
 \footnote{The action (\ref{sugra}) is presented in the ``string frame''.  
This terminology has to do with how the dilaton is separated from the metric.
 The ``Einstein frame'' is obtained by
absorbing a factor $e^{-\varphi(x)}$ into $g_{\mu\nu}(x)$. We shall not need to do this here.}  

The extremal black D3 brane solution of the super-gravity theory (\ref{sugra})  in the string frame is
the metric, dilaton and Ramond-Ramond 5-form fields
\begin{align}\label{metric}
&ds^2 = H(r)^{-\frac{1}{2}}\eta_{\mu\nu}dx^\mu dx^\nu +H(r)^{\frac{1}{2}}dx^m dx^m ~~,~~\\
&e^{2\varphi(r)}=g_s^2\\
&F_5 = (1+*)dx^0\wedge dx^1\wedge dx^2\wedge dx^3\wedge d\left(1-H^{-1}(r)\right)
\\ \nonumber &\mu,\nu=0,1,2,3~~;~~m,n=4,...,9 ~~,~~
r^2=x^mx^m\\ &H(r)=1+ \frac{L^4}{r^4}~,~~~
L^4=4\pi g_s N\alpha'^2
\label{L}
\end{align}
respectively. 
The element $\eta_{\mu\nu}dx^\mu dx^\nu$ is that of four-dimensional Minkowski space.
 $F_5=dA_4$ has $N$ units of flux through the 5-sphere, that is the 5-dimensional space
 with coordinates $x^4,...,x^9$ constrained by
 $(x^4)^2+...+(x^9)^2=1$.  This 5-sphere 
links the position of the black brane and the flux of the $5$-form integrated over the 5-sphere is $N$,
\begin{align}
\int_{S^5} F_{5}=N
\end{align}
This is an extremal charged black hole.  The event horizon is located at $r=0$. 

The function $H(r)$ is singular near the horizon at $r=0$.  However, the geometry is regular there, when $r<<L$, it approaches
\begin{align}
ds^2 &= \frac{r^2}{L^2}\eta_{\mu\nu}dx^\mu dx^\nu +\frac{L^2}{r^2}dx^m dx^m ~~,~~\nonumber \\
&=  \frac{r^2}{L^2}\eta_{\mu\nu}dx^\mu dx^\nu +\frac{L^2}{r^2}dr^2+L^2 d  \Omega_5^2
\end{align}
where  $ d  \Omega_5^2$ is the metric of the unit five sphere.   We shall sometimes write the 5-sphere coordinates more explicitly
as six-component unit vectors $\hat\theta$ where $\hat\theta^2=1$ and the line element is 
denoted $ d\hat\theta\cdot d\hat\theta$.  

What we obtain is the metric of the direct product spacetime $AdS_5\times S^5$.  $AdS_5$ is a space of constant negative curvature
whereas $S^5$ is a space of positive constant curvature and the radius of curvature of both spaces is $L$.    
The coordinate $r$ is the $AdS_5$ radius.  It has the units of length.   It is common to rescale $r$ so that it has energy units.  
This is done by redefining it as $r\to L^2 r$ whence
\begin{align}\label{ads}
ds^2
= L^2\left[  r^2\eta_{\mu\nu}dx^\mu dx^\nu +\frac{ dr^2}{r^2}+  d\hat\theta\cdot d\hat\theta\right]
\end{align}

Now, we must ask where the solution (\ref{metric})-(\ref{L}) is reliable as a background for the IIB string theory.   First of all, it is derived from the
supergavity equations of motion, without the corrections that should occur if these equations were embedded in string theory.  This
procedure will be self-consistent when those corrections are small.  They would be suppressed by powers of $\frac{\sqrt{\alpha'}}{L}$, and 
negligibly small  when the scale of the curvature of the space-time is much greater than the string scale, 
\begin{align}
L>>\sqrt{\alpha'}~~\to~~ 4\pi g_s N~>>~1
\end{align}
which is opposite to the limit of weakly interacting strings that we discussed in the previous section.  
This is the limit where the massive modes of the string decouple and the physics is governed by the 
lowest lying modes, the fields of ten dimensional type IIB supergravity. 

Once we have turned off the massive modes of the string, we have ten dimensional supergravity.  In principle,
this is a quantum field theory, including quantum gravity.  It is described by a classical solution only when the quantum
fluctuations are small.  If the classical solution is to be accurate,  we should  therefore shut off the quantum fluctuations of gravity. 
  This is done
by making the characteristic scale of the geometry, $L$,  much larger than the Planck length, 
\begin{align}\label{no_q_g}
L>>\ell_P
\end{align}
where the Planck length can be deduced from the coefficient of the action (\ref{sugra}), 
\begin{align}
\ell_P \sim\left( {\alpha'}^4g_s^2\right)^{\frac{1}{8}}
\end{align}
The action when evaluated on the classical black D3 brane is 
proportional to $\frac{L^8}{\ell_P^8}$
and it is large in this limit, justifying the semi-classical approximation.
The condition (\ref{no_q_g}) requires that
\begin{align}
N>>1
\end{align}
Thus, to be guaranteed to be a good description of the string theory, we need
$N$ to be large and the coupling to be strong, $4\pi g_s N>>1$.

Now, let us consider the low energy excitations of the IIB string about the background in equations 
(\ref{metric})-(\ref{L}). 
There are two ways in which a super-gravity excitation can have low energy.  First, it can have very long wave-length.  Long wavelength gravitons,
that is, gravitons with wavelengths much larger than $L$, and therefore with
the low energies, $E<< \frac{1}{L}$,   occupy the asymptotic region of the black D3 brane, where  $r>>L$.  
Their wavelengths are sufficiently long that they
do not fit into the throat of the geometry at smaller values of $r\sim L$.  What is more, their coupling to the other degrees of freedom scale to zero with
energy so that they are decoupled from those other degrees of freedom,  and from each other.  They are effectively
free gravitons propagating in flat ten-dimensional Minkowski space asymptotically far from the black hole. 

The other way that an excitation of the background can have low energy is to
occupy the region near the horizon, $r\sim 0$, that is, to have fallen down into 
the gravitational potential well which is the black hole.  
\footnote{For an extremal black hole, the horizon is a boundary of the space-time, rather than an event horizon.}
The energy of a string theory state in that region, that is, the energy that must be injected in order
to excite a typical string sitting at radius $r$ is $E_r\sim \frac{1}{\sqrt{\alpha'}}$, which is a high energy.  However,  as seen from an observer at infinity,
this energy 
is red-shifted to $\sqrt{-g_{00}} E_r=H(r)^{-\frac{1}{4}}E_r \sim 
\frac{r}{L}\frac{1}{\sqrt{\alpha'}}$ and this energy can be arbitrarily small if $r$ is small. 
Thus, within a distance $\sqrt{\alpha'}$ of the horizon, we can find all of the excitations of the string theory and they can
all be viewed by an observer at infinity as having low energy. 

The last statement, that we can find the full spectrum of the IIB string at small $r$ might worry the reader,
since we have said that the solution of super-gravity can only be trusted when the massive modes of the
string decouple.   This worry would be justified in that such a high energy excitation should back-react on
the geometry.   The answer to this question lies in the fact that the near-horizon geometry, with $AdS_5\times S^5$ metric
given in (\ref{ads}), is thought to be an exact solution of not only super-gravity, but the IIB string theory itself.  This means
that all corrections that would have arisen from integrating out massive string modes, which are proportional to $\frac{\alpha'}{L^2}$, 
should vanish in this region anyway.   It also means that the back-reaction even
from locally high energy string states vanishes.   

We conclude that the very low energy limit of the super-gravity degrees 
of freedom are of these two kinds, which are decoupled from each other, free super-gravitons
propagating in ten dimensional Minkowski space and the full IIB string theory in the 
near horizon geometry, $AdS_5\times S^5$.  

\subsection{The Maldacena conjecture}

Now, we return to the idea that, in the two scenarios described in the two previous sections, we are simply describing the same dynamical system
in two different limits, both where $N\to\infty$ holding $4\pi g_s N$ fixed and then, in the first where $4\pi g_s N<<1$ 
and the second where $4\pi g_s N>>1$ and in both cases at low energy $<<1/\sqrt{\alpha'}$. 
In both cases, there was a subsystem of almost free super-gravitons propagating on ten dimensional 
Minkowski space and we can identify these subsystems.  The great insight of Maldacena was to then identify the other 
two subsystems, the ${\mathcal N}=4$ theory of low energy open strings and the closed IIB string on the $AdS_5\times S^5$
near-horizon geometry.  The idea is that these are descriptions of one and the same system, each description
being  analytically tractable in a different limit.  The parameters are identified as
$g_{\rm YM}^2=4\pi g_s$, as in  equation (\ref{g_YM}).  Moreover,  defining $\lambda\equiv 4\pi g_s N=g_{\rm YM}^2N$, 
we have $L^2=\sqrt{\lambda}\alpha'$, as in equation (\ref{L}). It is in principle
a one-to-one identification of the two, including identification of all of the quantum states, observables,
operators and amplitudes.   As such, it is rather remarkable.  It says that one observer whose observational tools
are, for example, particle accelerators, might see this world as four space-time dimensional and containing
${\mathcal N}=4$ theory with gauge fields and adjoint representation
quarks, whereas a differently equipped observer would see exactly the same world as being ten dimensional and 
the elementary objects being strings.  Of course, $\lambda$ would be the same in both worlds, and if it were large,
for example, the second, string  observer would be able to do accurate calculations whereas the first, gauge field observer,
although he might have guessed that gauge fields where the degrees of freedom, 
would be frustrated to find that his theory to describe his world is strongly coupled. 

Given the duality between a gauge and string theory that we have been discussing, to make it
useful, we must identify what is dual to what.   One can go a long way in such an identification by
comparing the symmetries of the theories and identifying objects with the same quantum numbers.  Given
a set of gauge invariant local operators in the gauge theory, $O_i(x)$, the generating functional
\begin{align}
Z[\phi_0]~=~\left< e^{i\int d^4x \phi^i_0(x) O_i(x)}\right>
\end{align}
can be used to find correlation functions. One must take functional derivatives by the source fields $\phi^i_0(x)$ and
then set them to zero.   
In order to be gauge invariant, the operators $O_i(x)$ must be composite.  Moreover, it is convenient to 
 organize them according
to their scaling dimensions, $\Delta_i$, that is, they should obey
$$
-i\left[ D,O_i(x)\right]=i(x^\mu\partial_\mu+\Delta_i)O_i(x)
$$

On the string side of the correspondence, the procedure is as follows.   First, we identify
the degree of freedom of the IIB string which is dual to the operator $O_i(x)$ of the gauge theory.  Generally, the way that
degrees of freedom transform under symmetries as well as the quantum number $\Delta_i$ 
is a useful guide in this identification.  In practical terms, the degrees of freedom
are often fields which are in the string spectrum.  Then, for that degree
of freedom,  we require a boundary condition in the string theory on $AdS_5\times S^5$ that the field $\phi^i(x)$ approaches
a constant times the classical source function $r^{\Delta_i}\phi^i_0(x)$ at the boundary of $AdS_5$.   

The theory is then solved and the on-shell partition function of the
string theory is found, call it $Z_{\rm string}[\phi^i \to r^{\Delta_i}\phi^i_0]$, which is a functional of $\phi^i_0(x)$.  
The prescription for the generating functional is then

\begin{align}
\left< e^{i\int d^4x \phi^i_0(x) O_i(x)}\right>~=~Z_{\rm string}[\phi^i\to r^{\Delta_i} \phi^i_0]
\end{align}
In the parameter regime $\lambda>>1$ where the string theory is semi-classical and at large N where
it is super-gravity, the latter partition   function is gotten from the on-shell action in classical
super-gravity and 
 \begin{align}
\left< e^{i\int d^4x \phi^i_0(x) O_i(x)}\right>~\approx~e^{iS_{sugra}[\phi^i\to r^{\Delta_i} \phi^i_0]}
\end{align}
Here, $O^i(x)$ must the Yang-Mills theory operators which are dual to the fields of the supergravity theory.

\section{ ${\mathcal N}=4$ superconformal quantum field theory}

 In this section we shall outline some of the details of ${\mathcal N}=4$ supersymmetric Yang-Mills theory.  Much of
 this section is quite technical and it could be skipped on first reading of these notes.  Very few details beyond the 
 bosonic part of the Lagrangian density will be needed in the following sections. I include these details mainly for
 completeness. 
 
The ${\mathcal N}=4$ theory which appears on one side of the AdS/CFT duality is of interest for a number of reasons. 
Of course, it is the quantum field theory side of the example of the AdS/CFT correspondence that we have discussed in the 
previous section.    This duality has an
enormous number of potential applications in many contexts.
In addition, ${\mathcal N}=4$ is a simple, albeit nontrivial example of a supersymmetric and conformally symmetric 
quantum field theory in four dimensions where
the high degree of supersymmetry simplifies a number of properties but leaves the detailed dynamics of the theory
highly nontrivial.
 The conjectured integrability of the planar limit of the theory, and then also the IIB 
 string theory in $AdS_5\times S^5$ background in the limit of vanishing string coupling, has also been of great interest. For example, integrability
 can be used to do  computations of   the conformal dimensions of large classes of local operators with amazing efficiency. 
 
 The ${\mathcal N=4}$ theory is a highly non-trivial four dimensional quantum field theory which is thought to exhbit an
 exact  non-perturbative $SL(2,{\mathcal Z})$ duality, which interchanges electric and magnetic
fields and charges.  If we combine the theta-angle and coupling constant to form the complex
parameter
\begin{equation}\label{sl2z}
\tau = \frac{\theta}{2\pi}+\frac{4\pi i}{g_{YM}^2}
\end{equation}
the $SL(2,Z)$ has the generators
\begin{align}
\tau\to -\frac{1}{\tau}~~,~~\tau\to \tau+1
\end{align}
Finally, in the study of scattering amplitudes using modern techniques, 
an infrared cutoff ${\mathcal N}=4$ theory has been claimed to be the simplest interacting 
four dimensional Yang-Mills theory.

Historically, ${\mathcal N}=4$ theory  first appeared in the work of
Brink, Scherk and Schwarz \cite{Brink:1976bc}, who  constructed the four dimensional field theory
by using dimensional reduction of
 ${\mathcal N}=1$ supersymmetric Yang-Mills theory in ten dimensions.
The field content is
\begin{enumerate}
\item{}Six scalar fields $\phi^i(x)$, $i=1,\ldots,6$
\item{}A vector gauge field $A_\mu(x)$
\item{}Four Weyl spinor fields $\lambda_\alpha^a(x)$, $a=1,2,3,4$ and their conjugates $\bar\lambda_{\dot\alpha a}(x)$
\end{enumerate}
The sclar and vector fields are $N\times N$ Hermitian matrices and $\lambda$ and $\bar\lambda$ are Hermitialn
conjugates of each other. 
 Lorentz indices are raised and lowered by the Minkowski metric (\ref{minkowski_metric}). 
The spacetime vector indices are $\mu=0,1,2,3$, spinor indices are $\alpha,\dot\alpha=1,2$
and  $i=1,2,...,6$ and $a=1,2,3,4$ are $O(6)=SU(4)$ R-symmetry indices.
The un-dotted and dotted spinors transform under the $SL(2,C)$ universal cover of the Lorentz group in
the usual way.   The
spinor indices   are contracted using the
$SL(2,C)$ antisymmetric invariant tensors $\epsilon_{\alpha\beta}$ or $\epsilon^{\alpha\beta}$ and $\epsilon_{\dot\alpha\dot\beta}$ or $\epsilon^{\dot\alpha\dot\beta}$ where $-\epsilon_{21}=\epsilon_{12}=1=-\epsilon^{12}=\epsilon^{21}$.  Ergo $\lambda^\alpha \equiv \epsilon^{\alpha\beta}\lambda_\beta$, for example.
The sets of $2\times2$ matrices
$\sigma^\mu_{\alpha\dot\beta}$ and $\bar\sigma^\mu_{\dot\alpha\beta}$ are defined by the four dimensional Dirac matrices in a
representation where $\gamma^5$ is diagonal, 
\begin{align}
\gamma^\mu = \left[ \begin{matrix} 0 & \sigma^\mu \cr \bar\sigma^\mu & 0 \cr \end{matrix}\right]
\end{align}
and Dirac matrices obey the algebra
\begin{align}
\left[\gamma^\mu,\gamma^\nu\right]=-2\eta^{\mu\nu}
\end{align}
In a particular basis, $\sigma^\mu_{\alpha\dot\beta}=\left(-{\mathcal I},\vec\sigma \right)$ and
$\bar\sigma^\mu=\left( -{\mathcal I},-\vec\sigma\right)$ where $\vec\sigma$ are the Pauli matrices
\begin{align}
\sigma^1= \left[\begin{matrix}0&1\cr 1&0\cr\end{matrix}\right] ~,~
\sigma^2= \left[\begin{matrix}0&-i\cr i&0\cr\end{matrix}\right] ~,~
\sigma^3= \left[\begin{matrix}1&0\cr 0&-1\cr\end{matrix}\right]
\end{align}
Also,
\begin{align}
\sigma^{\mu\nu}_{\alpha\beta}=\sigma^\mu_{\alpha\dot\beta}\epsilon^{\dot\beta\dot\gamma}
\bar\sigma^\nu_{\dot\gamma\beta} - \sigma^\nu_{\alpha\dot\beta}\epsilon^{\dot\beta\dot\gamma}
\bar\sigma^\mu_{\dot\gamma\beta} \\
\bar\sigma^{\mu\nu}_{\dot\alpha\dot\beta}= \bar\sigma^\mu_{\dot\alpha\beta}\epsilon^{\beta\gamma}
\sigma^\nu_{\gamma\dot\beta} -\bar \sigma^\nu_{\dot\alpha\beta}\epsilon^{\beta\gamma}
\sigma^\mu_{\gamma\dot\beta}
\end{align}
For more details on these conventions, see \cite{Wess:1992cp}.

In component fields (this is the superfield action written in component fields in the  ``Wess-Zumino gauge''), the Lagrangian density of ${\mathcal N}=4$   is
\begin{align}\label{n=4Lagrangian}
{\mathcal L}&= {\rm Tr}\left[ -\frac{1}{2}F_{\mu\nu}F^{\mu\nu} 
-\sum_{i=1}^6D_\mu \phi^i D^\mu\phi^i 
+g_{\rm YM}^2\frac{1}{2}\sum_{i,j}\left[\phi^i,\phi^j\right]^2  \right. \nonumber
\\& \left. 
 -i\sum_{a=1}^4\bar\lambda^a\bar\sigma^\mu D_\mu\lambda_a
+g_{\rm YM}\sum_{iab}C_{i}^{ab}\epsilon_{\alpha\beta}\lambda^{\alpha }_a\left[\phi^i,\lambda^{\beta}_{ b}\right]
+g_{\rm YM}\sum_{iab}\bar C_{i ab}\epsilon^{\dot\alpha\dot\beta}
\bar\lambda_{\dot\alpha }^a\left[\phi^i,\bar\lambda_{\dot\beta }^b\right]\right]
\end{align}
where the trace is over the $N\times N$ matrix indices.
The first term is the usual Yang-Mills action and the second and third terms  contain the six adjoint representation scalar fields.
The matrices $C_{iab}$ and $\bar C_i^{ab}$ are obtained from the $8\otimes 8$ Euclidean Dirac matrices of six-dimensional Euclidean space,
in a chiral representation where $\gamma^1\gamma^2\ldots\gamma^6$ is diagonal, 
\begin{align}
\gamma_i = \left[ \begin{matrix} 0 & \bar C_{i\bar a\bar b} \cr   C_{i}^{ab} & 0 \cr \end{matrix} \right]~~,~i=1,...,6~,~a,b,\bar a,\bar b=1,2,3,4
\end{align}
normalized by the anti-commutator
\begin{align}
\left\{\gamma_i,\gamma_j\right\}=2\delta_{ij}
\end{align}
Shortly, we shall also need the spin tensors
 \begin{align}
{\bar C_{ijb}}^a=C_{i}^{ac}\bar C_{jcb}-C_{j}^{ac}\bar C_{icb} \\
{C_{ija}}^{b}=\bar C_{iac} C_{j}^{cb} - \bar C_{jac} C_{i}^{cb}
\end{align}

The dual of the field strength tensor is
\begin{align}
\tilde F_{\mu\nu}=\frac{1}{2}\epsilon_{\mu\nu\rho\sigma}F^{\rho\sigma}
\end{align}
where $\epsilon_{\mu\nu\rho\sigma}$ is the completely antisymmetric tensor with
$\epsilon_{0123}=1$.  The self-dual and anti self-dual field strength tensors are defined by
\begin{align}
F_{\mu\nu}^\pm = \frac{1}{\sqrt{2}}\left( F_{\mu\nu}\pm \tilde F_{\mu\nu}\right)
\end{align}

The supersymmetry transformations are generated by a Weyl spinor supercharge
\begin{align}
\delta_{\alpha}^{ a} X = \frac{1}{i}\left[ Q_{\alpha}^{ a} , X\right]_\mp
\end{align}
where the $\mp$ means that we should use a commutator or anti-commutator
if $X$ is bosonic or fermionic, respectively.
The transformations are
\begin{align}
\delta_{\alpha}^{ a}\phi^i &= C^{~iab}\lambda_{\alpha b} \\
\delta_{\alpha}^{ a}\lambda_{\beta b}& =~ F^+_{\mu\nu}~\sigma^{\mu\nu}_{\alpha\beta}~\delta_{~b}^{a}
~+~g_{\rm YM}\left[\phi^i,\phi^j\right]~(\bar C_{ij })^a_{~b} ~\epsilon_{\alpha\beta}  \\
\delta_{\alpha}^{ a} \bar\lambda_{\dot\beta}^{ b}&=C_{i}^{ab} \sigma^\mu_{\alpha\dot\beta} D_\mu\phi^i
\\
\delta_{\alpha }^{a} A^\mu &= \sigma^\mu_{\alpha\dot\gamma}\epsilon^{\dot\gamma\dot\beta}\bar\lambda_{\dot\beta}^{ a}
\end{align}

 The quantum field theory described by the Lagrangian (\ref{n=4Lagrangian}) contains no dimensional parameters and, at
 the classical level it has exact scale and therefore conformal invariance.  It also has the remarkable feature that the conformal
 invariance survives at the quantum level in that the beta-function for its only coupling constant $g_{\rm YM}$ vanishes.
 This is easily seen at the one-loop level by the old formula for the 1-loop beta function in Yang-Mills theory,
 \begin{align}\label{beta}
 \beta(g_{\rm YM},N)=-\frac{g_{\rm YM}^3}{16\pi^2}\left( \frac{11}{3}N-\frac{1}{6}\sum_k C_2({\rm scalars})-\frac{1}{3}
 \sum_k C_2({\rm Weyl~spinors})\right)
 \end{align}
 where $C_2$ is the quadratic Casimir invariant of the representation under which the scalar and spinor fields transform.
 $C_2=N$ for the adjoint representation.
 ${\mathcal N}=4$ has 6 real scalars and 8 Weyl spinors, so one can see that the right-hand-side of (\ref{beta})
 vanishes. The one-loop beta function is zero. 
  This was extended to two \cite{Tarasov:1976ef} and three loops  \cite{Avdeev:1980bh} \cite{Grisaru:1980nk}.  There is a general algebraic proof in the light-cone gauge \cite{Mandelstam:1982cb} as well as algebraic argument which we shall review here.

 \subsection{$PSU(2,2|4)$}

 The symmetries of ${\mathcal N}=4$ Yang Mills theory are described by Lie superalgebra
 $PSU(2,2|4)$.  In the following we shall describe the three components of this algebra, the Poincare
 algebra, its conformal extension, its supersymmetric extension and finally  its superconformal extension. 

\subsubsection{Poincare algebra}

 The Poincare symmetry of the field theory contains symmetry under space-time translations, rotations and Lorentz boosts.
 These transformations form a non-compact Lie group called the Poincare group.    
 The Lie algebra corresponding to the Poincare group
  is generated by four momentum operators $P_\mu$ which generate the spacetime
 translations and six (antisymmetric) $L_{\mu\nu}$ which generate  the rotations and Lorentz boosts. 
 In any relativistic quantum field theory, this Lie algebra should be realized by a set of Hermitian operators.
Moreover,  in a local
 quantum field theory, the appropriate Hermitian operators  are integrals
 of moments of components of a conserved, symmetric stress-energy tensor.  
 
 The commutation relations of the  Lie  algebra basis elements  are
 \begin{align}
 \left[ P_\mu,P_\nu\right]&=0 \label{PP}\\
 \left[ L_{\mu\nu},P_\lambda\right]& =-i\left(\eta_{\mu\lambda}P_\nu-\eta_{\nu\lambda}P_\mu\right)
 \label{MP}\\
 \left[ L_{\mu\nu} , L_{\rho\sigma} \right]& = -i\left( \eta_{\mu\rho}L_{\nu\sigma}+
 \eta_{\nu\sigma}L_{\nu\rho} - \eta_{\mu\sigma}L_{\nu\rho} - \eta_{\nu\rho}L_{\mu\sigma}\right)
 \label{MM}
 \end{align}
 The quantum field theory can be thought of as a set of infinite dimensional unitary representations of
 this algebra. 
 
\subsubsection{Conformal symmmetry}

According to the Coleman-Mandula theorem  \cite{Coleman:1967ad}, 
only certain extensions of the Poincare algebra are allowed  as the symmetry algebra of a quantum field theory that
has unitary time evolution. 
 One interesting possible extension of the Poincare algebra is the conformal algebra in four dimensions.  Field theories
 with this symmetry do not have asymptotic states (in fact the spectrum of the mass operator $P_\mu P^\mu$ is continuous)
 and they do not have an S-matrix.  However, they are of interest as generic renormalizable quantum field theories must 
  approach theories with conformal symmetry in their high and low energy limits.   The ${\mathcal N}=4$ Yang Mills theory
  will also have conformal symmetry.   The appropriate Lie algebra is that
 of the noncompact Lie group $SO(2,3)$ which has fifteen generators.  These include the Poincare generators with the
 algebra (\ref{PP})-(\ref{MM}),  as
 well as the  generator of dilations, $D$, and a generator of special conformal transformations, $K_\mu$, 
 so that, in addition to  (\ref{PP}-\ref{MM}),  the remaining
 algebra is
 \begin{align}
 \left[ D, P_\mu\right] &=-iP_\mu~,~
 \left[ D, L_{\mu\nu}\right]=0~,~
 \left[ D, K_\mu\right]=iK_\mu  \label{DK}\\
 \left[ L_{\mu\nu},K_\lambda\right]& =-i\left(\eta_{\mu\lambda}K_\nu-\eta_{\nu\lambda}K_\mu\right)\\
 \left[ P_{\mu},K_\nu\right]& =2i\left(L_{\mu\nu}-\eta_{\mu\nu}D\right)\\
 \left[ K_{\mu},K_\nu\right]& =0
 \end{align}
 An operator $O_\Delta(x)$ is said to have dimension $\Delta$ if it has the property
 \begin{equation}
 i\left[ D, O_\Delta(x)\right] =\left( x^\mu\partial_\mu +\Delta\right)O_\Delta(x)
 \label{dimension}
 \end{equation}
 Under a finite transformation, 
 \begin{equation}
 s^{iD}O_\Delta(x)s^{-iD}= s^{\Delta}O_\Delta(sx)
 \end{equation}
 Then, the algebraic relations in (\ref{DK}) imply that $K_\mu$ is a lowering operator for the dimension.
  To see this, consider the action of the special conformal generator $\left[K_\mu,O_\Delta(x)\right]$.  The dimension of the resulting operator
 is gotten by taking the commutator
 \begin{align}
 \left[ D, \left[K_\mu, O_\Delta(x) \right]\right]= -\left[ K_\mu, \left[ O_\Delta(x) ,D\right]\right]-\left[O_\Delta(x), \left[D,K_\mu \right]\right] \nonumber\\
 =  -i(\Delta-1)~\left[K_\mu.O_\Delta(x)\right]
 \end{align}
  where, we have used the Jacobi identity for commutators and the algebra in equations (\ref{dimension}) and
 (\ref{DK}). We see that the commutator with $K_\mu$ has lowered the dimension if $O_\Delta(x)$ by one unit. 
Similarly, $P_\mu$ is a raising operator,
  \begin{align}
 \left[ D, \left[P_\mu, O_\Delta(x) \right]\right]= -\left[ P_\mu, \left[ O_\Delta(x) ,D\right]\right]-\left[O_\Delta(x), \left[D,P_\mu \right]\right] \nonumber\\
 =  -i(\Delta+1)~\left[P_\mu.O_\Delta(x)\right]
 \end{align}

 In a relativistic quantum field theory, the dimensions of operators other than the identity must be positive and they have a lower
 bound.  If four dimensions, the lower bound for bosonic operators is $\Delta_{\rm min}=1$, the dimension of a free scalar field, 
 and for fermionic operators, it is $\Delta_{\rm min}=\frac{3}{2}$, the dimension of a free fermion field. 
 This means that, for any operator $O_\Delta(0)$ which has a given dimension, $\Delta$, the dimension cannot be lowered beyond a certain positive minimum value, that is, there must be operators such
 that
 \begin{equation}\label{primary}
 \left[ K_\mu , {\mathcal O}_{\Delta_0}(0)\right]=0
 \end{equation}
 Such an operator is called a {\it primary operator}.  It has dimension $\Delta_0$. 
 Commutators of a primary operator with $P_\mu$, 
  \begin{equation}
 \left[ P_{\mu_1}, \left[ P_{\mu_2},\ldots  \left[ P_{\mu_k}, {\mathcal O}_{\Delta_0}(0)\right]\ldots
 \right]\right]  =\left. (-i)^k\partial_{\mu_1}\partial_{\mu_2}\ldots\partial_{\mu_k} {\mathcal O}_{\Delta_0}(x)\right|_{x=0}
 \label{descendant}
 \end{equation} 
 raise the dimension of ${\mathcal O}_{\Delta_0}(0)$ from $\Delta_0$ to $\Delta_0+k$.  If $O_p(x)$ has dimension $\Delta_P$, then $\left[P_\mu,O_p(x)\right]$ has
 dimension $\Delta_p+1$, $\left[P_\mu,\left[ P_\nu, O_p(x)\right]\right]$ has dimension $\Delta_{p+2}$, and so on.
 The resulting operators are called {\it descendants}.  
  
 Note that we are evaluating $O_\Delta(x)$ at $x=0$, the
 fixed point of the conformal group. $O_\Delta(x)$ at other points is in principle gotten by considering the Taylor
 expansion using the descendants of $O_\Delta(0)$. 
 The set of all commutators of the generators of the conformal group with
 a primary operator forms a representation  of the conformal algebra which is infinite dimensional, as it must be
 for an non-compact Lie algebra.  Also, note that, since $L_{\mu\nu}$ commutes with $D$, it does not change the
 dimension of an operator.  It must therefore either commute with $O_\Delta(0)$, as would be the case for a scalar, 
 or there must be several $O_{\Delta s}(0)$ with the same dimension, so that
 \begin{equation}
 \left[ L_{\mu\nu},O_{\Delta s}(0)\right] = iO_{\Delta s'}(0)[\Sigma_{\mu\nu}]_{~ s}^{s'}
 \end{equation}
 that is, the operators $O_{\Delta s}$ carry a finite dimensional representation
 of the algebra with Lorentz spin matrices $\Sigma_{\mu\nu}$ 
 

 Conformal symmetry determines the coordinate dependence of two-point functions up to normalization, which
 can be chosen so that
 \begin{align}
  \left< O_{\Delta_1}(x) O_{\Delta_2}(0)\right>= \frac{ \delta_{\Delta_1,\Delta_2}}{|x|^{\Delta_1+\Delta_2}}
  \end{align}
 The three-point functions are also determined, up to a constant,
\begin{align}
  \left< O_{\Delta_1}(x) O_{\Delta_2}(y)O_{\Delta_3}(z)\right>= \frac{c(\Delta_1,\Delta_2,\Delta_3)}
  {|x-y|^{\Delta_1+\Delta_2-\Delta_3}|y-z|^{\Delta_2+\Delta_3-\Delta_1}|z-x|^{\Delta_3+\Delta_1-\Delta_2}}
  \end{align}
  The fact that these are determined follows from the ability to use finite conformal transformations to
  move the two or three points to some reference values.  This no longer works for higher point functions
  since the coordinates have some invariant combinations, for four-point functions
  one can construct two combinations of coordinates, called harmonic ratios, which are invariant under conformal
  transformations.
  
\subsubsection{ Supersymmetry}

 Another allowed extension of the Poincare algebra is obtained by adding some anti-commuting generators, which
 transform like spinors under Lorentz transformations and rotations.   Consider the
 un-dotted spinor supercharges $Q_{\alpha}^{ a}$ and their conjugate dotted spinors
 $\bar Q_{\dot\alpha a}$ where $a=1,...,{\mathcal N}$ in theories with different degrees of supersymmetry, 
Generally, ${\mathcal N}=1,2,4$.  
Here we will consider the case specific to  ${\mathcal N}=4$.  
In addition to the Poincare algebra (\ref{PP})-(\ref{MM}), a supersymmetric relativistic quantum field theory
contains  $Q_{\alpha}^{ a}$ and  
 $\bar Q_{\dot\alpha a}$ which have the  
 algebra
 \begin{align}
 \left[ P_\mu,Q_{\alpha}^{ a}\right]&=0 ~,~\left[ P_\mu,\bar Q_{\dot\alpha a}\right] =0\\
 \left[ L^{\mu\nu}, Q_{\alpha}^{ a}\right]&= i\sigma^{\mu\nu}_{\alpha\beta}\epsilon^{\beta\gamma}Q_{\gamma}^{ a} ~,~
 \left[ L^{\mu\nu},\bar Q_{\dot\alpha a}\right]= i\bar\sigma^{\mu\nu}_{\dot\alpha\dot\beta}
 \epsilon^{\dot\beta\dot\gamma}\bar Q_{\dot\gamma}^{ a} \\
\{Q_{\alpha}^{ a},Q_{\beta}^{ b}\}& =0~,~\{Q_{\alpha}^{ a},\bar Q_{\dot\beta b}\}=2\delta^a_b\sigma^\mu_{\alpha\dot \beta}P_\mu~,~
\{\bar Q_{\dot\alpha  a},\bar Q_{\dot\beta  b}\}=0 \label{QQ}
\end{align}
The indices $a,b=1,2,3,4$ transform under $SU(4)$ R-symmetry. To distinguish the supercharges $Q_{\alpha}^{ a}$ and
$\bar Q_{\dot\alpha a}$ from the conformal supercharges which we shall introduce below, these are called the {Poincare
supercharges}.

\subsubsection{ Superconformal algebra}

Now, let us consider the situation where a supersymmetric theory also happens to be a conformal field theory.
It is clear from the algebra (\ref{QQ}) that the supercharges have dimension $\frac{1}{2}$,
\begin{align}
\left[ D, Q_{\alpha}^{ a}\right] = -\frac{i}{2}Q_{\alpha}^{ a} ~,~
\left[ D, \bar Q_{\dot\alpha a}\right] = -\frac{i}{2}\bar Q_{\dot\alpha a}
\end{align}
To close the algebra, more anti-commuting generators are needed.  These are the conformal supercharges, $S_{\alpha a}$ and $\bar S_{\dot\alpha }^{a}$,
which are commutators of the Poincare supercharges with $K_\mu$,
\begin{align}
\left[ K^\mu, Q_{\alpha}^{ a} \right]& = \sigma^\mu_{\alpha\dot\alpha}\epsilon^{\dot\alpha\dot\beta}\bar S_{\dot\beta}^{ a}
~,~
\left[ K^\mu,\bar Q_{\dot\alpha a} \right] = \bar\sigma^\mu_{\dot\alpha\alpha}\epsilon^{\alpha\beta}S_{\beta a}   \\
\left[ D, S_{\alpha a}\right] &= \frac{i}{2}S_{\alpha a} ~,~
\left[ D, \bar S_{\dot\alpha}^{ a}\right] = \frac{i}{2}\bar S_{\dot\alpha}^{ a}
\end{align}
The remainder of the algebra is
\begin{align}
\{S_{\alpha a},S_{\beta b}\}& =0~,~\{S_{\alpha a},\bar S_{\dot\beta}^{ b}\}=2 \delta^b_a\sigma^\mu_{\alpha\dot \beta}K_\mu~,~
\{\bar S_{\dot\alpha}^{ a},\bar S_{\dot\beta}^{ b}\}=0 \label{SS}  \\
\{Q_{\alpha}^{ a}, S_{\beta b}\}&= \epsilon_{\alpha\beta} \left[
  { {{C  }_{ij} }^a }_bT^{ij}+  \delta_{\,\, b}^aD\right] +\frac{1}{2}\delta_{\,\, b}^a
\sigma_{\alpha\beta}^{\mu\nu}L_{\mu\nu} \label{QS}\\
\{\bar Q_{\dot\alpha a}, \bar S_{\dot\beta}^{ b}\}&= \epsilon_{\dot\alpha\dot\beta}\left[  {\bar C_{ij a} }^{~\,\,\,\,b} T^{ij}+\delta^{\,\,b}_aD
\right] +\frac{1}{2}\delta^{\,\, b}_a
\bar\sigma_{\dot\alpha\dot\beta}^{\mu\nu}L_{\mu\nu} \label{QS2}\\
\{Q_{\alpha}^{ a}, \bar S_{\dot \beta}^{ b}\}&=~0 ~,~\{\bar Q_{\dot\alpha a}, S_{\beta b}\}=0
\end{align}
where $T^{ij}$ is a generator of the $SO(6)\sim SU(4)$ Lie algebra.  
In SO(6) it generates a rotation in the $ij$-plane and the spin matrices $C$ and $\bar C$ convert them to the 4 and $\bar 4$ representations
of $SU(4)$, so that 
$$
[ {\bar C_{ij a} }^{~\,\,\,\,b} T^{ij},Q_\alpha^c]= i\delta_a^{\,\, c} Q_\alpha^b~,~[ {\bar C_{ij a} }^{~\,\,\,\,b} T^{ij},\bar Q_{\dot\alpha c}]=
-i  \bar Q_{\dot\alpha a}  \delta^b_{\,\, c}
 $$
A primary operator $  {\mathcal O}_{\Delta_0}(0)$ is an operator which has a conformal dimension $\Delta_0$
and which obeys $\left[ K_\mu,   {\mathcal O}_{\Delta_0}(0)\right]=0$. 

We note that (anti-)commutators of an operator of dimension $\Delta$
with $S$ and $\bar S$, if they are nonzero, must yield operators with dimensions $\Delta - \frac{1}{2}$.  We can therefore
use $S$ and $\bar S$ to lower the dimension of an operator.  
This process of lowering the dimension using $S$ and $\bar S$ must truncate at operators of a minimal dimension. 
There must therefore exist operators
which obey
\begin{align}\label{conformalprimary}
\left[ S_{\alpha}^a, \tilde {\mathcal O}_{\Delta_0 s}(0)\right]=0
~,~
\left[ \bar S_{\dot\alpha a},\tilde{\mathcal O}_{\Delta_0 s}(0)\right]=0
\end{align}
(these are anticommutators if $\tilde {\mathcal O}_{\Delta_0s}(0)$ is fermionic).   An operator $\tilde {\mathcal O}_{\Delta_0s}(0)$ 
with this property is 
called a {\it superconformal primary operator}.  
Here, we have anticipated that there can be more than one such operator and the label $s$
runs over the set of operators.  

It is easy to see that superconformal primary operators are always primary operators.  
However, the converse does not have to be true, it is possible that a primary  is not a superconformal primary. 

Commutator brackets of the other algebra elements with a superconformal primary generate a representation of $PSU(2,2|4)$ (modulo the usual
need to eliminate null vectors). Generally, representations of the superconformal algebra can split into several representations of the conformal subalgebra.

\subsubsection{ Short Representations}

 Though all nontrivial unitary representations of the superconformal algebra are infinite dimensional, there is a sense in
  which some have fewer states than others.  Such representations are called {\it short representations}.  They occur 
  when the superconformal primary operator, as well as obeying
  $
\left[ S_{\alpha a}, \chi_{\Delta s k}(0)\right]=0
$ and $
\left[ \bar S_{\dot\alpha}^{ a},\chi_{\Delta s k}(0)\right]=0
$, also commutes with some of the Poincare supercharges,
\begin{align}\label{short}
\left[ Q_{\alpha}^{ a} , \chi_{\Delta s k}(0)\right]=0~{\rm for~some~of~the}~a
\end{align}
There are called {\it chiral primary operators} and we have denoted them by the symbol
$\chi_{\Delta s k}(0)$.   It has conformal dimension $\Delta$
and, for clarity, we have displayed its indices $s$ and $k$ explicitly.  These  indices  transform under spin and internal symmetry, 
respectively.  

Assume that a chiral primary operator exists and  consider the values of $a$ for which (\ref{short}) holds.  
Then,  trivially, the following double commutator vanishes identically,
\begin{align}
\left\{S_{\beta b},~ \left[Q_{\alpha}^{ a} , \chi_{\Delta s k}(0)\right]\right\}=0
\end{align}
Using the Jacobi identity, we find
\begin{align}
0= \{S_{\beta b} ,[ Q_{\alpha}^{ a} , \chi_{\Delta s k}(0)]\}=-\{ Q_{\alpha a} ,[\chi_{\Delta s k}(0), S_\beta^b]\}
-[ \chi_{\Delta s k}(0) ,\{S_\beta^b, Q_{\alpha a} \}] \nonumber \\
=[~\chi_{\Delta s k}(0) ~,~ \epsilon_{\alpha\beta}\left[ C^{ija}_{b}T_{ij}+\delta_b^aD\right]+2\delta_b^a
\sigma_{\alpha\beta}^{\mu\nu}M_{\mu\nu}~ ]
\label{shortidentity}\end{align}
where we have used the fact that $\chi_{\Delta s k}(0)$ is a super-conformal primary (\ref{conformalprimary})
and  the super-conformal algebra equation (\ref{QS}). Now, the fact that both $L_{\mu\nu}$ and $T_{ij}$ commute with $D$ tells us that the set
of operators degenerate with $\chi_{\Delta s k}(0)$ must carry a representation of both $SO(1,3)$ and $SO(6)$.
This implies that
\begin{align}
[ T_{ij},\chi_{\Delta s k}(0) ] = i\chi_{\Delta s k'}(0)[t_{ij}]_{\,\,k}^{k'} ~~,~~[L_{\mu\nu} , \chi_{\Delta s k}(0) ]=i \chi_{\Delta s' k}(0)[\Sigma_{\mu\nu}]_{\,\, s}^{s'}
\end{align}
where $t_{ij}$ and $\Sigma_{\mu\nu}$ are representations of $SU(4)\sim SO(6)$ and $SO(3,1)$ respectively.  Equation
(\ref{shortidentity}) then implies
\begin{align}
 \epsilon_{ \alpha \beta}\bar C^{ija}_{~~~~b}[t_{ij}]_k^{k'}\chi_{\Delta s k'}(0)- \epsilon_{\alpha\beta}\delta_a^b
\Delta \chi_{\Delta s k}(0)-\delta^a_b
 \sigma_{ \alpha \beta}^{\mu\nu}[\Sigma_{\mu\nu}]_s^{s'}\chi_{\Delta s' k}(0) =0
\label{shortrequirement}
\end{align}
For a Lorentz scalar, $\Sigma_{\mu\nu}=0$.  Let us consider that
case.  We must understand the implication of (\ref{shortrequirement}).
$C^{ij}$ are a set of matrices in the fundamental representation of $SU(4)$ which is
a rank 3 algebra.  Assuming a certain normalization (and remembering that rotations in orthogonal planes in $R^6$ should
commute), we can write the Cartan subalgebra
as
\begin{align}\small
C^{12}=\left[ \begin{matrix} 1 & 0 & 0 & 0 \cr 0&1&0&0\cr 0&0&-1&0\cr 0&0&0&-1\cr \end{matrix}\right] ~,~
C^{34}=\left[ \begin{matrix} 1 & 0 & 0 & 0 \cr 0&-1&0&0\cr 0&0&1&0\cr 0&0&0&-1\cr \end{matrix}\right] ~,~
C^{56}=\left[ \begin{matrix} 1 & 0 & 0 & 0 \cr 0&-1&0&0\cr 0&0&-1&0\cr 0&0&0&1\cr \end{matrix}\right]
\end{align}
The remaining 12 generators have vanishing diagonal elements. The eigenvalues of these matrices are
the weights of the fundamental representation.  If the eigenvalues of $t_{12},t_{34},t_{56}$ are $(J_1,J_2,J_3)$, respectively, the 1-1 component of (\ref{shortrequirement}) is then $J_1+J_2+J_3=\Delta$, and the 2-2 component is $J_1-J_2-J_3=\Delta$.
The equation is obeyed when  $J_2=J_3=0$ and when $J_1=\Delta$. The chiral primary operator has $\Delta=J_1$ which, due to the Lie algebra
which quantizes $J_1$,  must be an integer.   More importantly, it is independent of the coupling constant $g_{\rm YM}$ or the number $N$.  
Since it is not expected to change discontinuously as $g_{\rm YM}$ is varied, we expect that it is independent of the coupling and it is given
by its value at $g_{\rm YM}=0$, that is, by the free field lime, which is its engineering dimension. 

An operator of this form is found by taking the complex combination of scalar fields $z=\phi^1+i\phi^2$, and the generator whose eigenvalue is $J_1$
generates a phase transform of $z=\phi^1+i\phi^2$.   The composite 
\begin{align}
O_{k}(0)={\rm Tr} z(0)^k
\end{align}
$\left[ T_{12},O_k\right]=kO_k$, $\left[T_{34},O_k\right]=0$, $\left[T_{56},O_k\right]=0$.  This operator has tree level dimension $k$.  
However, its dimension is
determined by the algebra, so it must also have dimension $k$, unaffected by radiative corrections in the
interacting theory.

This, a super-conformal primary operator which commutes with half of the supercharges. It is a called a {\it chiral
primary operator}.  A single trace operator of this form is
\begin{align}
O_k~=~{\rm Tr} \phi^{(i_1}(0)\phi^{i_2}(0)\ldots \phi^{i_k)}(0) ~-~{\rm Traces}
\end{align}
which is made completely symmetric and traceless in its indices.

  This is the first no-renormalization theorem -- the dimension of chiral primary operators does not renormalize.
  It implies that, suitably normalizable, their two-point functions are known exactly,
  \begin{align}
  \left< O_{\Delta_1}(x) O_{\Delta_2}(0)\right>= \frac{ \delta_{\Delta_1,\Delta_2}}{|x|^{\Delta_1+\Delta_2}}
  \end{align}
  (where $\Delta$ are positive integers).
  There is a actually a stronger statement that can be made. The three-point functions of these operators
is also though to not renormalize:
\begin{align}
  \left< O_{\Delta_1}(x) O_{\Delta_2}(y)O_{\Delta_3}(z)\right>= \frac{c(\Delta_1,\Delta_2,\Delta_3)}
  {|x-y|^{\Delta_1+\Delta_2-\Delta_3}|y-z|^{\Delta_2+\Delta_3-\Delta_1}|z-x|^{\Delta_3+\Delta_1-\Delta_2}}
  \end{align}
  and the coefficients $c(\Delta_1,\Delta_2,\Delta_3)$ are given by their tree-level values.

  It is also an important fact that all of the operators in
  the entire representation of the superconformal algebra that is generated
  from a chiral primary operator has the same anomalous dimension.
  An of such an operator is, for example ${\rm Tr}F_{\mu\nu}F^{\mu\nu}$ itself that
  can be obtained from ${\rm Tr}\phi^i\phi^j-\frac{1}{6}\delta^{ij}{\rm Tr}(\phi^k)^2$
  by taking four commutators with supercharges.   This implies that the operator has
  dimension four.  Since the Lagrangian also must have dimension 4, the coupling constant $g_{YM}$
  must be dimensionless and therefore scale independent.  A similar argument for conformal invariance
  was originally given in \cite{Sohnius:1981sn}.

\section{ Holographic Wilson loops}

In this section, I will give a simple review of the computation of the Wilson loop, both in the Yang-Mills theory and on the gravity
side of the AdS/CFT duality. We will not give a complete review of this subject, instead we shall concentrate on explaining some of the
basic ideas.   

As we have already discussed, the Wilson loop is the quantum expectation value of the holonomy of a heavy
quark wave-function when   we would drag the quark around a closed curve $C$ in spacetime (which 
we shall take to be Euclidean).  The loops that are readily accessible to us in AdS/CFT are rather special ones which contain the scalar fields in the exponential
\begin{align}
W[C]=<{\rm Tr}{\mathcal P}e^{\int_0^1d\tau\left( i\dot x^\mu(\tau)A_\mu(x(\tau))+|\dot x(\tau)|\hat\theta\cdot \phi(x(\tau))\right)}>
\end{align}
Here, the expectation value is in the vacuum state of the  ${\mathcal N}=4$ super-conformal Yang-Mills theory with $SU(N)$ gauge group. 
 The functions $x^\mu(\tau)$ sweeps out the curve $C$ as $\tau $ goes from $0$ to $1$. The scalar  is multiplied by $\hat\theta$ which is a unit vector residing on the $S^5$.  This expectation value is the measure of the holonomy of 
the wave-function of a heavy quark when it is parallel transported along the curve $C$.   

A weak coupling expansion of the Wilson loop begins with a Taylor expansion of the path ordered exponential  
\begin{align}\nonumber
W[C]=\left[ N
+\frac{g_{\rm YM}^2}{2}\int d\tau \int d\tau\left[| \dot x(\tau)||\dot x(\tau')|\hat\theta^i\hat\theta^j <{\mathcal P}\phi^{ab}_i(x(\tau))
\phi^{ba}_j(x(\tau'))>_{0}\right.  \right.
\\
\left.\left. -\dot x^\mu(\tau)\dot x^\nu(\tau') <  {\mathcal P}A^{ab}_\mu(x(\tau))A^{ba}_\nu(x(\tau'))>_{0}\right]+\ldots\right]
\end{align}
where, in the Feynman gauge,  the free-field propagators are proportional to Green functions for the four-dimensional Laplacian,  
$$
<A^{ab}_\mu(x)A_\nu^{cd}(y)>_0 = \frac{\delta_{\mu\nu}\delta^{ad}\delta^{bc}}{8\pi^2(x-y)^2}~,~
<\phi^{ab}_i(x)\phi_j^{cd}(y)>_0 = \frac{\delta_{ij}\delta^{ad}\delta^{bc}}{8\pi^2(x-y)^2}
$$
We have taken the gauge group to be $U(N)$.  

We get the expression
\begin{align}\label{wc}
W[C]={\rm Tr}\left[ 1
+\frac{\lambda}{16\pi^2}\int d\tau d\tau' \frac{ |\dot x(\tau)||\dot x(\tau')|-\dot x(\tau)\cdot\dot x(\tau') }{(x(\tau)-x(\tau'))^2}+\ldots\right]
\end{align}
The contributions from the vector and scalar fields in this expression have the interesting consequence that the singularity at 
$\tau  \to \tau'$ that would normally lead to a linearly ultraviolet divergence
is absent here.  
For smooth trajectories, the integrals in (\ref{wc}) are finite.

\subsection{Straight line and circle Wilson loop}

The result (\ref{wc}) is the leading terms in an expansion of the Wilson loop in the coupling constant  $\lambda$. 
The integrals over the curve parameter can easily be done for some simple examples.  For example, for a straight line Wilson loop,  where
$x^\mu(\tau)=(\tau,0,0,0)$ the correction term vanishes and 
\begin{align}
W[{\rm straight ~ line}]=N
\end{align}

  For a circle on the other hand, $x^\mu(\tau)=a(\cos\tau,\sin\tau,0,0)$, the 
result is 
\begin{align}\label{perturbative_circle}
W[{\rm circle}]= N\left( 1+\lambda/8+\ldots\right)
\end{align}
Scale invariance tells us that it does not depend on the radius of the circle.  This is an expansion in small $\lambda$ about $\lambda=0$. 

The prescription for computing this Wilson loop at strong coupling is to seek a minimal surface $D$ whose boundary traces the curve $C$
placed on the boundary of $AdS_5$ (at $r\to\infty$),  while sitting at a point on $S_5$. The Wilson loop expectation value is the classical limit of the
disc amplitude which is given by 
\begin{align}\label{holwl}
W[C]=\frac{1}{g_s} \exp\left( -\frac{1}{2\pi\alpha'} \inf_{D:\partial D=C}A[D] +M   L[C]\right)
\end{align}
where  $A[D]$ is the area of $D$. The counter-term, $M L[C]$, where $ L[C]$ is the length of $C$ and $M$ is the infinite mass of the heavy quark,
 cancels a linear divergence that always occurs in the first term in the exponent.   The factor of $\frac{1}{g_s}= \frac{4\pi}{\lambda}N$ arises from the
 fact that this is a  disc amplitude.  The factor of $N$ is expected, on the gauge theory side, it comes from the trace which is taken to find the Wilson loop.  
 The remaining factor $ \frac{4\pi}{\lambda}$ should be regarded as sub-leading in the
 semi-classical limit.
 
 As an example, consider the Wilson loop which is a straight line, $x^\mu(\tau)=(\tau,0,0,0)$.   By symmetry, one might expect that 
 the world-sheet which extremises the area is simply the surface which extends along the $x^1$-direction and the $r$-direction, with the
 embedding equations 
 $$x^\mu(\tau,\sigma)=(\tau,0,0,0)$$ $$r(\tau,\sigma)=\sigma$$
 and the position on the 5-sphere is given by a unit vector is 6-dimensional space $\hat\theta$,
 so that the intrinsic metric is 
 $$
 ds^2 = L^2\left[\frac{d\sigma^2}{\sigma^2}+\sigma^2d\tau^2\right]
 $$
 which is the metric of $AdS_2$ with the same radius of curvature, $L$, as the $AdS_5$ that it is embedded in. 
 The area of this world-sheet is given by 
$$
\frac{1}{2\pi\alpha'}A[D] = \frac{1}{2\pi\alpha'} \int d\tau \int_0^\infty d\sigma \sqrt{\det g} = \frac{1}{2\pi\alpha'} \int d\tau \int_0^\infty d\sigma L^2
$$
This expression is infinite.  To make sense of it, we could call the factor $$\int d\tau = L[C]$$ the length of the curve. 
The other factor is 
\begin{align}
\label{mass}
\frac{1}{2\pi\alpha'} \int_0^\infty d\sigma L^2 = \frac{\sqrt{\lambda}}{2\pi}\int_0^\infty dr\equiv M
\end{align}
 which we
shall show can be interpreted as the mass of the heavy quark. This also coincides with quark masses in probe brane constructions of heavy
quarks \cite{karch}.   Then, the area term in equation (\ref{holwl}) and the length term
cancel each other
$$
\left[\frac{1}{2\pi\alpha'}A[D] -ML[C]\right]_{\rm straight_line}=0
$$
 We  find that, for a straight line, 
$W=\frac{1}{g_s}$ which, to within the accuracy of our computation, is the same as $W=N$ (where we  recall that $N=\frac{1}{g_s}\cdot \frac{\lambda}{4\pi}$.  This is identical to the result (\ref{perturbative_circle}) which we found at weak coupling, but of course here
it is good in the strong coupling regime.  
We might conjecture that this is so in between these two limits, that when $C$ is an infinite
straight line,  $W[C]=N$ to all orders
in perturbation theory.  This has indeed been confirmed at weak coupling to a few orders at small $\lambda$.   
Improving the string theory calculation to check the higher orders or the strong coupling expansion in $1/\sqrt{\lambda}$, and in particular, to check
that the coefficient is indeed changed from $ \frac{1}{g_s} $ to $N $
seems to be more difficult.  It would require solving for the fluctuations of the full type IIB superstring sigma model about the classical solution.  
At the time of writing of these notes, this has not been done in full detail.   Some discussion can be found in references \cite{Aguilera-Damia:2018twq}
and \cite{Aguilera-Damia:2018bam}.  

 A reason for this triviality of the straight line would be supersymmetry.  The straight-line Wilson loop operator of interest commutes with half of the Poincare supercharges of the ${\mathcal N}=4$ theory. 
 
 For the circle Wilson loop, $x^\mu(\tau)$ is, for example, the locus of $(x^1)^2+(x^2)^2=a^2$.   The minimal surface in $AdS_5$ with this boundary, and with maximal symmetry, turns out to be the locus of $(x^1)^2+(x^2)^2+\frac{1}{r^2}=a^2$, which is a Euclidean $AdS_2$ embedded in $AdS_5$.  Using a particular world-sheet coordinate
 system, the embedding equations  are
 \begin{align}\label{embedding}
 x^1(\sigma,\tau) =\frac{a}{\cosh a\sigma}\cos a\tau
~,~ x^2(\sigma,\tau)=\frac{a}{\cosh a\sigma }\sin a\tau
~,~ r(\sigma,\tau) = \frac{1}{a}\coth a\sigma~,~\hat\theta\subset S^5
\end{align}
Here, $\sigma=0$ is located at $r=\infty$  and $\sigma=\infty$ is located at $r=\frac{1}{a}$. 
The induced metric of the world-sheet is that of a Euclidean $AdS_2$ black hole
\begin{align}\label{ads2blackhole}
ds^2 =L^2\left(\frac{a}{\sinh a\sigma}\right)^2 \left[ d\sigma^2+d\tau^2   \right]
\end{align}
where the horizon located at $\sigma=\infty$. The absence of a conical singularity at the poiint $\sigma=\infty$ requires peridicity of the Euclidean time with a particular
period.  In that region,
$$
ds^2 \approx L^2  \left(4a^2e^{-2a\sigma} d\sigma^2+4a^2e^{-2a\sigma}d\tau^2 \right)=4L^2\left[dy^2+a^2y^2d\tau^2\right]
$$  and there is no conical singularity at $y=0$ if the time has the correct periodic identification,  $\tau\sim\tau+2\pi/a$ (which we already knew from 
equation (\ref{embedding})).   Of course, in Euclidean quantum field theory, peroidic time means finite temperature, which, in this case, is interpreted as a Hawking temperature of the black hole. 
The Hawking temperature is thus 
\begin{align}\label{Hawking_temperature}
T_H = \frac{a}{2\pi}
\end{align}
  For the surface which we have been discussing, we can evaluate the exponent in the contribution to the Wilson loop in (\ref{holwl}), 
$$
\frac{1}{2\pi\alpha'}A[D] -M   L[C] =\frac{1}{2\pi\alpha'}\int_0^{2\pi/a }d\tau \int_0^\infty d\sigma L^2\left(\frac{a}{\sinh a\sigma}\right)^2
-\left( \frac{\sqrt{\lambda}}{2\pi}\int_0^\infty dr\right)~ (2\pi a )
$$
$$
= \frac{\sqrt{\lambda}}{2\pi}(2\pi a) \left[ \int_0^\infty   d\sigma\frac{dr}{d\sigma}
- \int_0^\infty dr\right]
 =-  \sqrt{\lambda} a\int_0^{\frac{1}{a} }  dr
  =-\sqrt{\lambda}
$$
where, in the last term in the first line,  we have used the length of the curve, $L[C]=2\pi a$ and the heavy quark mass given in equation (\ref{mass}).  
We see that the circle Wilson loop,
in the limit of large $\lambda$ is then given by 

\begin{align}\label{strong_circle}
W[{\rm circle}]=Ne^{\sqrt{\lambda}}
\end{align}
Generally, because of conformal symmetry, we expect that, the circle Wilson loop is independent of the radius of the circle.  Indeed, our results
indicate that,  in the large $N$ limit,  
\begin{align}\label{matrixmodel3}
W[{\rm circle}]=Ne^{f(\lambda)}, ~~~~f(\lambda)= \left\{\begin{matrix}  \frac{\lambda}{8} & \lambda<<1 \cr  \sqrt{\lambda} & \lambda>>1\cr \end{matrix}\right.
\end{align}
In fact, this function is known to all orders in $\lambda$.  It was conjectured using re-summed perturbation theory \cite{Erickson:2000af}  
and confirmed using supersymmetric localization \cite{Pestun:2007rz}
that the circle Wilson loop is given by the matrix model
 \begin{align}\label{matrixmodel0}
 W[ {\rm circle}]=  \frac{   \int dM   ~e^{-\frac{8\pi^2N}{\lambda}{\rm Tr}M^2 }~~ {\rm Tr}e^{M}  }
  {   \int dM   ~e^{-\frac{8\pi^2N}{\lambda}{\rm Tr}M^2 }  }
   \end{align}
   This integral can be computed exactly \cite{Drukker:2000rr}
   \begin{align}\label{matrixmodel1}
 W[ {\rm circle}] =  L^1_{N-1}[-\lambda/4N]e^{\lambda/8N}
   \end{align}
   where 
   \begin{align}
   L^n_m =  \frac{1}{n!} e^x x^{-m}\left(\frac{d}{dx}\right)^n  x^{m+n}e^{-x}
   \label{laguerre}
   \end{align}
    is the Laguerre polynomial.  
   The  large $N$ limit is 
     \begin{align}\label{matrixmodel2}
\lim_{N\to\infty} W[ {\rm circle}] =  N\frac{  2}{\sqrt{\lambda}}I_1(\sqrt{\lambda}) 
   \end{align}
   where $I_k$ is the modified Bessel function.   Asymptotic expansions of the Bessel function can be used to show 
   that the small and large $\lambda$ limits agree with our computations summarized in (\ref{matrixmodel3}).

 \subsection{Wilson loops and heavy quarks}
 
It is enlightening to put the discussion of the Wilson loop into a more physical context.  
To do this, we begin with ${\mathcal N}=4$ super-conformal  Yang-Mills theory with gauge group $SU(N+1)$.  
Consider the scenario where  the scalar fields gets a non-zero  vacuum expectation value, 
$$
\left< \phi^{N+1~N+1}_6 (x)\right> =\varphi 
$$
The expectation values of all other components of that and the other scalar fields vanish, 
$$
\left< \phi^{ab}_6(x)\right> =0 ~{\rm either}~a\neq N+1{\rm ~or~}b\neq N+1
$$
$$
\left< \phi^{a,b}_i(x)\right> =0~,~i=1,2,...,5
$$
 With such a condensate, the gauge symmetry will be reduced by the Higgs mechanism, from  $SU(N+1)$ to $SU(N)\times U(1)$ and 
 the SO(6) R-symmetry is reduced to SO(5). 
 Once the symmetry is broken, there are three types of fields.  There is still a full ${\mathcal N}=4$ super-multiplet of massless fields 
 with gauge group $SU(N)$ and which,
 in isolation, would be a super-conformal quantum field theory. These massless fields are singlets under the $U(1)$ gauge symmetry.
  Then there are massless fields which constitute a full ${\mathcal N}=4$ super-multiplet with gauge group $U(1)$ and 
  which are singlets under the SU(N) gauge symmetry. 
 Then, there is a short $\frac{1}{2}$-BPS super-multiplet of massive
 $W$-fields.  The latter multiplet contains vector, scalar and spinor fields (which we generically refer to as $W$-fields).  These
 transform under the fundamental representations of both the residual $SU(N)$ and $U(1)$.  
  The bosonic fields
in the super-multiplet are  vectors
$$
W_\mu^a=A_\mu^{aN+1}~~,~~\bar W_\mu^a=A_\mu^{N+1a}
$$
 and scalars
 $$
\Phi_i=\phi_i^{aN+1}~~,~~\bar\Phi_i=\phi_i^{N+1a}~~,~~I=1,2,...,5
$$
These fields get masses from the commutator terms in the Yang-Mills action, for example, if we write
$$
\phi_i^{ab}~\to~\varphi\delta_{i6}\delta^{aN+1}\delta ^{bN+1} ~+~\phi_i^{ab}
$$
Then
$$
 -g_{\rm YM}^2 \sum_{i<j=1}^6 \left[\phi_i(x), \phi_j(x)\right]^2 ~\to~
  g_{\rm YM}^2\varphi^2\sum_{i=1}^5 \sum_{a=1}^N \bar\Phi_i^a\Phi_i^a+{\rm ~cubic~}+{\rm quartic}
$$ 
we would identify the mass as $M_W=g_{\rm YM}|\varphi|$. 
Also, the SO(6) R-symmetry is reduced to SO(5) symmetry by the choice of which scalar gets a vacuum expectation value.
The orientation of this expectation value is generally a point on the coset $SO(6)/SO(5)=S^5$ which can be represented by a six-dimensional
unit vector $\hat\theta_0$.   In the above discussion, we have taken the simple case where $\hat\theta_0$ is in the 6-direction.

Let us consider the quantum amplitude whose modulus squared is the probability
that one of the scalar fields in the $W$-boson super-multiplet 
will propagate from position $x_i^\mu$ with color index $a$  to position $x_f^\mu$ with color index $b$.  
(The ${\mathcal N}=4$ theory is not in a confining phase, so global color should be a good quantum number and we can ascribe 
``colors'' to particles.)   
 Let us denote the amplitude by 
$
{\mathcal A}_{ab}(x_f,x_i) 
$.  
We discuss $
{\mathcal A}_{ab}(x_f,x_i) 
$  in  the large $N$ limit, at  both the weak coupling and the strong coupling limits, in both
cases at the limit where the $W$-boson mass is large.  
 For this purpose, it is instructive to use a world-line functional integral
representation of the   propagator on Euclidean space
%
\begin{align}\label{amplitude}
{\mathcal A}_{ab}(x_f,x_i) 
=\int_0^\infty [dT] \int[dx^\mu(\tau)] e^{-S[x]} \left<  \left[ {\mathcal P}e^{\oint d\tau(i\dot x^\mu(\tau) A_\mu(x(\tau))+|\dot x(\tau)|\hat\theta_0\cdot \vec\Phi(x(\tau))}\right]_{ab} \right>
\end{align}
Of course, this propagator is only non-zero in the gauge-fixed theory and even then it can depend on the way in which the
gauge is fixed.  Nevertheless, it should contain some physical information, such as the value of the physical mass of the
W-boson.  This could be deduced from the asymptotic behaviour for large proper time, where
\begin{align}\label{amplitude1}
{\mathcal A}_{ab}(x_f,x_i)  \sim  \delta_{ab}   \exp\left(-M\sqrt{(x_f-x_i)^2}\right)
\end{align}
(Remember that we are in Euclidean space.  If we were in Minkowski space, the decaying exponential
would be replaced by a phase $e^{-iM\sqrt{-(x_f-x_i)^2}}$.)
That the coefficient $M$ of the proper time in the exponent in equation (\ref{amplitude1})  is gauge invariant is equivalent to the statement that the pole in the propagator is gauge invariant. 

The residual global colour symmetry of the gauge fixed theory tells us that equation (\ref{amplitude1})  is equivalent to 
\begin{align}\label{amplitude}
{\mathcal A}_{ab}(x_f,x_i) 
=\frac{\delta_{ab}}{N}\int_0^\infty [dT] \int[dx^\mu(\tau)] e^{-S[x]} \left< {\rm Tr} {\mathcal P}e^{i\oint(\dot A+\Phi^1)}  \right>
\\
=\frac{\delta_{ab}}{N}\int_0^\infty [dT] \int[dx^\mu(\tau)] e^{-\tilde S[x]} 
\end{align}
\begin{align}
\tilde S=\int_0^1 d\tau \left[\frac{1}{4T}\dot x^\mu(\tau)\dot x^\mu(\tau)+M^2 T\right] -\ln W[C]
\end{align}
where $W[C]= \left< {\rm Tr} {\mathcal P}e^{i\oint(\dot A+\Phi^1)}  \right>$ is the open Wilson loop  
for the curve $C$ which is parameterized by $x^\mu(\tau)$.
The potential term in the action is given by the logarithm of this Wilson loop, 
and  the functional integration has  boundary conditions, $x(1)=x_f$ and $x(0)=x_i$. 

The Wilson loop computes the interaction of the
particle with the remaining massless  fields of the ${\mathcal N}=4$ supersymmetric Yang-Mills theory
which are by themselves an ${\mathcal N}=4$ theory with residual $SU(N)$ gauge group.  
The influence on the amplitude (\ref{amplitude}) of the W-bosons as well as the residual $U(1)$ gauge fields and other fields in their super-multiplets
can be neglected in the large $N$ limit.  This is due to the fact that they transform in the fundamental or singlet representations of $SU(N)$,
respectively, and their loops  therefore contribute only to sub-leading orders in the large $N$ limit.  The one, open W-boson line which is allowed in the amplitude is the
one that is computed by the world-line functional integral in (\ref{amplitude}). 

We have  written a gauge fixed version of the world-line functional integral where $T$ is not the einbein, as it would be in the
completely time reparametrization invariant version of the integral, but here it is a constant which plays the role of Schwinger's proper time.
We have used $[dT]$ to denote the measure whose details depend on the regularization of the functional integral over the $x^\mu(\tau)$. The $T$-dependence
of the measure can be found by comparing with the Schwinger representation for the propagator of a free particle where
\begin{align}\label{amplitude2}
{\mathcal A}_0(x_f,x_i) 
=\int_0^\infty \frac{ dT  }{16\pi^2 T^2}~ e^{-\frac{(x_f-x_i)^2}{4T}-M^2T} =\frac{M\sqrt{(x_f-x_i)^2}K_1(M\sqrt{(x_f-x_i)^2})}
{4\pi^2(x_f-x_i)^2}
\end{align}
A derivation of the world line path integral for the scalar field propagator using 
zeta-function regularization can be found in the appendices of reference \cite{Gordon:2014aba}.

When the mass of the W-boson is large compared to any other quantity with dimensions (here, these could only by functions of the trajectory itself),
we can study this propagator in the semiclassical approximation.  The dimensionless number that controls this approximation is $\frac{1}{M\sqrt{(x_f-x_i)^2}}$.
To study the semiclassical approximation, we begin with the classical equations of motion 
\begin{align}\label{classical_e_o_m}
-\frac{\ddot x^\mu(\tau)}{2T}\left. -\frac{\delta}{\delta\tilde  x^\mu(\tau)}\ln W[\tilde x]\right|_{{\tilde x}^\mu(\tau)=x^\mu(\tau)}=0~,~ T^2=\frac{1}{4M^2}\int_0^1 d\tau (\dot x^\mu)^2
\end{align}
where we see that the presence of the Wilson loop creates a force term, the analog of the Lorentz force on a relativistic charged particle.  
This term should be evaluated on the trajectory $x^\mu(\tau)$.   The solution of equations (\ref{classical_e_o_m}) is a straight line trajectory,
\begin{align}\label{classical_solution}
x^\mu_0(\tau)= (x^\mu_f-x^\mu_i)\tau + x^\mu_i~,~T=\frac{1}{2M}\sqrt{(x_f-x_i)^2}
\end{align}
To see this, we note that $$-\frac{1}{2T} \ddot x_0^\mu(\tau) =0$$  Moreover, we can conjecture that
the functional derivative of the Wilson loop vanishes when it is evaluated on an infinite straight line
\begin{align}\label{lorentz_force1}
\left.  \frac{\delta}{\delta x^\nu(\tau)} \ln W[x^\mu]\right|_{x^\mu={\rm straight~line}}=0
\end{align}
This is indeed true of the perturbative expression given in equation (\ref{wc}), even for a finite straight line.  Indeed, for the infinite straight line, it is known that
the first derivative by the contour, evaluated on the straight line, can be expressed as an anticommutator of a supercharge with another fermionic
operator \cite{Semenoff:2004qr}.  
Therefore, as long as the superconformal symmetry of the $SU(N)$ ${\mathcal N}=4$ theory remains intact, equation
(\ref{lorentz_force1}) must hold, independent of the size of $\lambda$.    

We conclude that the straight line is a legitimate saddle point of the functional integral in equation (\ref{amplitude}).   
 We can proceed to evaluate the action on that saddle point.  
The world-line action contributes the term 
$$
\int_0^1 d\tau \left[\frac{1}{4T}\dot x^\mu(\tau)\dot x^\mu(\tau)+M^2 T\right] = M\sqrt{(x_f-x_i)^2}
$$
The semi-classical limit of the amplitude is thus given by plugging the classical solutions (\ref{classical_solution}) into the action, to get
\begin{align}\label{formprop} 
{\mathcal A}_{ab}(x_f,x_i) 
 \sim e^{-M\sqrt{(x_f-x_i)^2}} ~\delta_{ab} 
\end{align}

Let us review what we have learned from this computation.  Our only inputs were large $N$, which allowed us to ignore loops of the W-bosons, and the large mass limit for the W-boson, which allows us to do a semi-classical computation.  Then, in addition, we know that (\ref{lorentz_force1}) is independent of the coupling constant, and, in fact, independent of the large $N$ expansion. This tells us that the straight line is a solution of the classical equation (\ref{classical_e_o_m}).  Moreover, since the Wilson line cannot correct the exponential behaviour -- it does not depend on $M$ -- and, because of conformal symmetry it does not depend on $(x_f-x_i)^2$ either. 
This tells us that $M$ is not renormalized and that the semi-classical large N limit must always be of the form (\ref{formprop}).

 Now, let us move to the strong coupling side, which is described by the IIB string theory on $AdS_5\times S^5$background.
 On that side, we must first find the appropriate state of the string theory --  the one that is dual to the state of ${\mathcal N}=4$ theory
 with $SU(N+1)$ gauge symmetry spontaneously broken to $SU(N)\times U(1)$ by the Higgs mechanism.
 
 The  $AdS^5\times S^5$  background of the string theory is the gravitational field (in the near horizon region) due to a stack of $N+1$ D3 branes.  
 We create the state with $SU(N+1)$ Higgsed to $SU(N)\times U(1)$ by  separating one of the D3 branes from the stack of $N+1$ D3 branes
 and putting it at some distance from the stack. There are six directions in which they can go in being taken away from the stack and remain parallel to
 the branes that are in the stack.  We will call the direction that is chosen the unit 6-vector $\hat\theta_0$.  
 
  There are then three types of open strings.  There are those which begin and end on the remaining
 stack of $N$ D3 branes, those which begin and end on the separated D3 brane and those which stretch from the stack of $N$ D3 branes to the
 separated D3 brane.  
 
 The strings with connect the stack of N D3 branes to the separated D3 brane have a mass gap which is  proportional to the 
 distance of separation times the string tension.   The lowest energy states of these strings are the massive $W$-boson super-multiplet.  The lowest modes of open strings with both ends attached to the stack of $N$ D3 branes or both ends attached to the separated D3 brane are two ${\mathcal N}=4$ super-multiplets, the first with $SU(N)$ gauge symmetry and the second with $U(1)$ gauge symmetry.  In the limits which create the $AdS_5\times S^5$ background, the separated brane is located at a constant value of the AdS radius, $r=r_M$, and it is extended in the $x^\mu$ directions, parallel to the horizon at $r=0$ and the boundary at $r=\infty$ and it sits at a point on $S^5$
 given by the unit 6-vector $\hat \theta_0$,  the direction in which the D3 brane was separated from the stack. 
 
 It is interesting that the separated D3 brane will float at a fixed radius in $AdS^5$.   This is a consequence of the fact that it is a $\frac{1}{2}$-BPS state
 of the string theory where the energy of the state does not depend on the distance of separation.  Moving the D3 brane along the radius is a flat direction in the energy landscape.   The D3 brane is attracted to the stack of D3 branes by the gravitational interaction.  This attraction is exactly balanced by a repulsive interaction due to their Ramond-Ramond charges so that the net force is equal to zero.   Another way to see this is to consider the separated D3 brane a probe brane.  Then, in the leading order in string perturbation theory, the disc amplitude, the free energy of the separated D3 brane
 is given by the Dirac-Born-Infeld action augmented by a Chern-Simons term for the Ramond-Ramond field
 $$
 S_{\rm DBI}=T_3\int d^4x\left[ -\sqrt{-\det(g_{\mu\nu}-2\pi\alpha'F_{\mu\nu})}+\omega^{(4)}\right]
 $$
For a flat brane located at radius $r_M$ the embedding metric gotten by setting $r=r_M$ and leaving only $dx^\mu$ nonzero
 in the $AdS_5$ metric to get the induced metric of the D3 brane
 $$
 g^{D3}_{\mu\nu}dx^\mu dx^\nu = L^2 r_M^2 dx_\mu dx^\mu
 $$
The Ramond-Ramond 4-form  $\omega^{(4)}$ given by 
 $$
 \omega^{(4)}= {L^4}{r_M^4}dx^0  dx^1 dx^2 dx^3
 $$
If (for future reference) the brane world-volume gauge field has a constant electric field, $F_{01}=E$,
  \begin{align}\label{d3_dbi}
 S_{\rm DBI}=\frac{L^4r_M^4}{(2\pi)^3{\alpha'}^2 g_s}\int d^4x\left[ -\sqrt{1-\frac{(2\pi{\alpha'})^2}{L^4r_M^4}E^2 }+1\right]
 \end{align}
 We see that, when $E=0$, the action (\ref{d3_dbi}) is zero, independent of the radius $r_M$. When $E\neq 0$, but is small,  and using
 $4\pi g_s=g_{\rm YM}^2$, the energy of
 the electric field is
 $$
 S_{\rm DBI}\approx \frac{1}{g_{\rm YM}^2}E^2
 $$
 which is what we expect for the energy of the $U(1)$ field (up to a factor of 4 which has been absorbed into the 
 normalization of $g_{\rm YM}^2$).

 Thus we describe the appropriate state of the string theory as a probe D3 brane floating in $AdS^5$, parallel to both
 the boundary and the horizon, at radius $r_M$. Since, in this strong coupling limit,  $\sqrt{\lambda}>>1$, we can treat the string theory semi-classically. 
 We also work at weak string coupling $g_s$ where we only need to compute the disc amplitude.  We therefore take the semiclassical
 limit of the disc amplitude which is  the semi-classical limit of
 \begin{align}\label{string_amplitude}
 {\mathcal A}_{ab}(x_f,x_i)=\frac{\delta_{ab}}{N}~\frac{1}{g_s} \int[dXdrd\hat\theta \ldots]e^{-\frac{L^2}{4\pi\alpha'}  \int \left( r^2\partial X\cdot\bar\partial X + \frac{1}{r^2}
 \partial r\bar\partial r+\partial\hat\theta\bar\partial\hat\theta\right)+\ldots}
 \end{align}
 where $\partial = \frac{\partial}{\partial(\sigma+i\tau)}$ and 
 we have written only the bosonic part of the sigma model action as the Polyakov action in the conformal gauge. 
 The other fields in the supersymmetric
 world-sheet theory do not contribute to the leading order in the large $\sqrt{\lambda}$ limit.  
The factor $\frac{\delta_{ab}}{N}$ needs some explanation.  The initial state in the amplitude is an open string which is suspended between the
separated D3 brane and the D3 brane   labeled ``$a$'' in the stack of $N$ D3 branes and located at $x^\mu=x_i^\mu$ and at the point $\hat\theta=\hat\theta_0$ on
the 5-sphere which coincides with the direction in the 6-dimensional space in which the separated D3 brane was pulled away from the stack. 
The final state in the amplitude is an open string which is suspended between the
separated D3 brane and the D3 brane   labeled ``$b$'' in the stack and located at $x^\mu=x_f^\mu$ and $\hat\theta=\hat\theta_0$. 
The disc-like world-sheet should interpolate between these initial and final strings.   This is only possible if the endpoint of the string
follows the same D3 brane, therefore the amplitude must vanish unless $a=b$, hus the $\delta_{ab}$.
 Then, by symmetry,  the amplitude is independent from the index $a$ so  that we can average over index, thus the $\frac{1}{N}$.
 Then the stack of D3 branes is replaced by the
$AdS_5\times S^5$ geometry, and the information about individual D3 branes in the stack is lost, the open string must simply go to the
 $AdS_5$ horizon at $r=0$. 
 
 In the large $\sqrt{\lambda}$ limit, to implement the semi-classical approximation, 
 we look for a solution of the classical field equations for the embedding functions of the string,
 $$\left(
 r(\sigma,\tau), x^\mu(\sigma,\tau), \hat\theta(\sigma,\tau)\right)$$ into the $AdS_5\times S^5$ spacetime.  These equations are
 obtained from the action in the functional integral in equation (\ref{string_amplitude}) by a variational method.  They are
$$
-\partial_a\left(\frac{1}{r^2}\partial_a r\right)+r\partial_a X\partial_aX-\frac{1}{r^3}\partial_a r\partial_a r=0
$$
$$
\partial_a(r^2\partial_a X_\mu )=0
$$
$$
\partial_a^2\hat\theta-\hat\theta (\hat\theta\cdot\partial_a^2\hat\theta) =0~~,~~\hat\theta^2=1
$$
with the appropriate boundary conditions.
These are solved  by
$$
r(\sigma,\tau)=\frac{1}{ \sqrt{(x_f-x_i)^2}~\sigma}
~~,~~
X(\sigma,\tau)=(x_f-x_i)\tau+x_i~~,~~\hat\theta(\sigma,\tau)=\hat\theta_0
$$
This ansatz also solves the Virasoro constraints, 
 $$
  r^2\partial_\sigma X\cdot\partial_\tau X + \frac{1}{r^2}
 \partial_\sigma r\partial_\tau r +\partial_\sigma\hat\theta\cdot\partial_\tau\hat\theta=0$$
 $$
  r^2 \partial_\sigma X\cdot\partial_\sigma  X - r^2\partial X_\tau\cdot\ \partial_\tau X-\frac{1}{r^2}\partial_\tau r\partial_\tau r+ \frac{1}{r^2}
  \partial_\sigma r \partial_\sigma r+\partial_\sigma\hat\theta\cdot\partial_\sigma\hat\theta-\partial_\tau\hat\theta\cdot\partial_\tau\hat\theta=0
 $$
 which are associated with fixing the conformal gauge for the worldsheet coordinates in the functional integral in equation (\ref{string_amplitude}). 
 In order to satisfy the boundary conditions, $\sigma$ should be integrated from $\frac{1}{ \sqrt{(x_f-x_i)^2}~r_M}$ to $\infty$ and $\tau$ from $0$ to $1$.
 The on-shell action is
 $$
 \frac{\sqrt{\lambda}\alpha' }{4\pi\alpha'}  \int \left( r^2\partial X\cdot\bar\partial X + \frac{1}{r^2}
 \partial r\bar\partial r+\partial\hat\theta\cdot\bar\partial\hat\theta \right)= \frac{\sqrt{\lambda}}{2\pi }   \sqrt{(x_f-x_i)^2}~r_M$$
 where we have used $L^2=\sqrt{\lambda}\alpha'$
 and the amplitude is 
 \begin{align}\label{formprop1} 
 {\mathcal A}(x_f,x_i)_{ab}\sim \frac{\delta_{ab}}{N}\frac{1}{g_s}e^{- \frac{\sqrt{\lambda}~r_M}{2\pi}   \sqrt{(x_f-x_i)^2}}
 \end{align}
  This result is interesting.  It has the form given in equation (\ref{formprop} ) above, however, unlike (\ref{formprop} ) which is valid where 
  $M\sqrt{(x_f-x_i)^2}>>1$, (\ref{formprop1}) is
  valid in  when $\sqrt{\lambda}>>1$.  If we assume that they are sumultaneously valid,  
we can make the identification 
\begin{align}\label{Wmass}
M = \frac{\sqrt{\lambda} r_M}{2\pi }~~,~~r_M = \frac{2\pi M}{\sqrt{\lambda}}
\end{align}
The result is also compatible with the Wilson loop at strong coupling being equal to one.  This is notwithstanding the factor
$\frac{1}{N}\frac{1}{g_s}=\frac{4\pi}{\lambda}$ which should be computed by fluctuations about the classical solution.

 The other Wilson loop which is sometimes computable is the circle and we can develop something similar to the above reasoning for that case too.
 This has been done in an application to the study of the Schwinger pair production by a strong electric field where circle like instantons dominate the
 semiclassical limit of the Euclidean world-line path integral that is used to compute the imaginary part of the vacuum energy \cite{Semenoff:2011ng}. In the following we will
 review another development which is related to the circle, but it is in Lorentzian space-time.  As the reader will see, this subject is incomplete and there are
 several interesting open directions. 
 
 Consider the following thought experiment.   We consider the same scenario as above, Yang-Mills theory with $SU(N)$ gauge group
 broken by the Higgs mechanism to $SU(N)\times U(1)$.  We will work in Lorentzian space-time with real, rather than Euclidean time.  
 We will use the residual $U(1)$ symmetry as one which leads to what we will call a conserved electric charge.  The only objects in the theory
 which carry this electric charge are the $W$-bosons. We will consider the situation where a constant electric field which couples to this charge permeates
 the whole of space-time.   This will be a weak field, sufficiently weak that we can neglect Schwinger pair production of charged particle-antiparticle pairs which should
 always occur at a slow rate in a constant electric field.   The criterion for neglecting
 pair production  of particles of mass $M$ in and electric field of strength $E$  is 
 $\frac{E}{M^2}<<1$.  In the string theory, it was argued in reference \cite{Semenoff:2011ng} that, to avoid Schwinger pair production, we must
 also require that $\frac{\sqrt{\lambda}}{2\pi}\frac{E}{M^2}<<1$.   Then, we will consider a scalar particle in the $W$-boson super-multiplet and we will imagine that we prepare a state where we inject the particle from spatial infinity, travelling at almost the speed of light, and in a direction which is precisely anti-parallel to the electric field. The particle, when injected, has a particular
 color state with label $a$.  
We will ask what the quantum amplitude is for the particle to return to precisely the same position, with precisely the oppositely directed velocity after a time which is 
precisely equal to its classical travel time. We will consider the limiting case where the initial velocity of the particle approaches the speed of light, the injection position
is infinitely far from the place where the particle stops and turns around and the travel time approaches infinity.   Some of the ideas which we discuss here appeared in references \cite{Hubeny:2014zna}
and \cite{Hubeny:2014kma}.  
 
 The classical trajectory of a relativistic particle in a constant electric field has constant proper acceleration.  It 
is therefore a Rindler trajectory.    Our first aim is to do a semi-classical computation of the amplitude
   \begin{align}
  {\mathcal A}_{ab}  = \frac{\delta_{ab}}{N}\int[dx][dT] e^{iS}\left<{\rm Tr}{\mathcal P}e^{i\oint d\tau\left( \dot x^\mu(\tau)A_\mu(x(\tau))+\Phi_i(x(\tau)\hat\theta^i_0\right)}\right>
  \end{align}
  in the context of the gauge theory, 
  where the world-line action of the particle is
  $$
S=\int d\tau \left[\frac{1}{4T}
\dot x_\mu(\tau)\dot x^\mu(\tau) -M^2T +\frac{E}{2}(x_0\dot x_1 -x_1\dot x_0) \right]
$$
The first terms are  the same  as we used for the relativistic scalar particle in the previous discussion.  The terms with coeficient $E$ are
the coupling to the electric field,  $\int d\tau \dot x^\mu(\tau)A_\mu(x(\tau))$ where we use the gauge $A_\mu = -\frac{1}{2}F_{\mu\nu}x^\nu$
for a constant electromagnetic field $F^{\mu\nu}$, and then specialize to constant electric field. 
The equations of motion for the integration variables, $(x_\mu(\tau), T)$ are
\begin{align}
&-\frac{1}{2T}\ddot x_0(\tau)+E\dot x_1(\tau)+\frac{1}{i}\frac{\delta}{\delta x^0(\tau)}\ln W[C]=0\label{emo1}\\
&-\frac{1}{2T}\ddot x_1(\tau)+E\dot x_0(\tau)+\frac{1}{i}\frac{\delta}{\delta x_1(\tau)}\ln W[C]=0\\
&-\frac{1}{2T}\ddot x_{2,3}(\tau)-\frac{\delta}{\delta x_{2,3}(\tau)}\ln W[C]=0\\
&T^2=-\frac{1}{4M^2}\int_{-\tau_P/2}^{\tau_P/2}d\tau\dot x_\mu(\tau)\dot x^\mu(\tau)\label{emo4}
\end{align}
Then, we can use symmetry to argue that, when evaluated on the Rindler trajectory, $\frac{\delta}{\delta x^\mu(\tau)}\ln W[C]$ vanishes. 
As a consequence, equations (\ref{emo1})-(\ref{emo4}) are solved by the Rindler trajectory itself, with constant proper acceleration $\frac{E}{M}$ due
to the electric field, and in which we use the boundary conditions of our
thought experiment to write  the solution of equation (\ref{emo1})-(\ref{emo4})  as
\begin{align}
&x_\mu(\tau) = \left(  \frac{M}{ E} \sinh   \frac{E}{M}\tau, \frac{M}{ E} \cosh   \frac{E}{M} \tau ,0,0\right)\label{trajectory}\\
&T=\frac{\tau_P}{2M}~,~  -\tau_P/2\leq \tau\leq \tau_P/2
\end{align}
Here, $\tau_P$ is the total proper time accumulated by the W-boson during its flight and the initial and final points are
\begin{align}
&(x_f)_\mu= \left(  \frac{M}{ E} \sinh   \frac{E}{2M}\tau_P , \frac{M}{ E} \cosh   \frac{E}{2M}  \tau_P  ,0,0\right)\nonumber\\
&(x_i)_\mu = \left( - \frac{M}{ E} \sinh   \frac{E}{2M} \tau_P , \frac{M}{ E} \cosh   \frac{E}{2M}  \tau_P  ,0,0\right)\nonumber
\end{align}
The initial and final speeds, $v_{f,i}=\tanh   \frac{E}{2M}  \tau_P $ should be close to the speed of light, $v_{f,i}\sim 1$.  Together with our previous
weak field assumption, we need to be in the regime
$$1<<  \frac{M^2}{E}<< M\tau_P $$  
which we shall assume from now on. 
Due to gauge invariance issues, we shall consider only the part of the phase of the amplitude ${\mathcal A}$ which
grows linearly in $\tau_P$.  The coefficient of $\tau_P$ should be related to the energy of the particle in its rest frame. 
In fact, let us evaluate the world-line action on the classical trajectory that we have found.  The result is
\begin{align}\label{onshellaction}
S=-M\tau_P+\frac{M}{2}\tau_P
\end{align}
and, in this limit
\begin{align}
\label{amp1}
  {\mathcal A}_{ab}  \approx \frac{\delta_{ab}}{N} e^{-iM\tau_P+iM\tau_P/2}~W[{\rm Rindler}] 
\end{align}
We have separated the on-shell action (\ref{onshellaction}) into two parts in order to remind ourselves that the first term is the  energy 
of the $W$ boson in its rest frame, equal to its rest mass, $M$, times $\tau_P $, as we would expect.  The second term in (\ref{onshellaction}), $\frac{M}{2}\tau_P$, 
 is from the
interaction energy of the W-boson with the electric field, which cancels exactly half of its rest energy. The on-shell action becomes the 
phase in the amplitude (\ref{amp1}).  The remaining
factor is the open Wilson loop evaluated on the Rindler trajectory (\ref{trajectory}).  In the case of the straight line we argued that, as a consequence
of symmetry, the Wilson loop could not contribute to the phase.   We shall see that this is not the case here.  This is due to the presence of two dimensionful
parameters, the proper time $\tau_P$ and the acceleration $\frac{E}{M}$.   

The Rindler trajectory in equation (\ref{trajectory}) is a hyperbola,  the locus of the equation
$$
(x_0)^2-(x_1)^2+\frac{M^2}{E^2}=0
$$
where, $\frac{E}{M}$ is the proper acceleration.  This is the Lorentzian analog of the circle which would be gotten by putting $x_0\to -ix_0$.  
We therefore expect that some of the simplifications that characterize the circle Wilson loop should also apply to this trajectory.  Indeed, this is the case.
For example, the sum of the vector and the scalar particle propagators connecting any two points on the trajectory is a constant,
Using
$$
\left<{\mathcal T}A^{ab}_\mu(x )A^{cd}_\nu(y)\right>_0 = \frac{  \delta^{ad}\delta^{bc} \eta_{\mu\nu} }{ 8\pi^2(x-y)^\mu(x-y)_\mu +i\epsilon}
$$
$$
\left<{\mathcal T}\Phi^{ab}_i(x )\Phi^{cd}_j(y)\right>_0 = \frac{  \delta^{ad}\delta^{bc} \delta_{ij} }{ 8\pi^2(x-y)^\mu(x-y)_\mu +i\epsilon}
$$
(where the subscript 0 on the bracket indicates that it should be evaluated in the gauge fixed ${\mathcal N}=4$ theory with coupling $g_{\rm YM}$ set
to zero), 
we find, upon plugging in the Rindler trajectory (\ref{trajectory}), 
$$
\dot x^\mu(\tau)\dot x^\nu(\tau')\left<{\mathcal T}A^{ab}_\mu(x(\tau))A^{cd}_\nu(x(\tau'))\right>_0+\sqrt{-\dot x(\tau)^2}\sqrt{-\dot x(\tau')^2}\left<
{\mathcal T}\Phi^{ab}(x(\tau))\cdot\hat\theta_0 \Phi^{cd}(x(\tau'))\cdot\hat\theta_0\right>_0
$$
$$
=\frac{1-\cosh\frac{E}{M}(\tau-\tau')} { 8\pi^2\frac{M^2}{E^2}(2-2\cosh(\frac{E}{M}(\tau-\tau'))-i\epsilon} \delta^{ad}\delta^{bc} = \frac{E^2}{16\pi^2 M^2} \delta^{ad}\delta^{bc} 
$$
  This is similar
to the circle Wilson loop, however, in the case of the circle, the constant propagator connects points on a finite, circle contour, where as here the contour
is open and of proper length $\tau_P$, which we shall eventually take as being large.  Each integral over
the contour parameter produces  a factor of $\tau_P$.  Let us
consider the leading order which is similar to the Euclidean case, given in equations (\ref{wc}) and (\ref{perturbative_circle}), 
\begin{align}\label{rind1}
W[{\rm Rindler}] = N\left[   1-\frac{\lambda}{8}\left( \frac{E}{2\pi M}\right)^2 \tau_p^2+\ldots \right]
\end{align}
The first thing that we learn from equation (\ref{rind1}) is  that the correction is large and it diverges as $\tau_P$ becomes large. 
Perturbation theory only makes sense if this correction is small. We shall therefore keep $\tau_P$ finite and assume that $\lambda$ is
small enough that the corrections are indeed small.  We will resum some of the perturbation theory and, afterward, we will examine the result
by relaxing the assumption of small corrections.    To sum all of the Feynman diagrams which contain only lines whose endpoints are on the trajectory, so
that their propagators are constants, all we need to do is to solve the combinatorics of the matrix indices.   As in the case of the Euclidean circle, this 
is easiest done using a matrix model.  If these are the only contributions that we keep, we compute 
\begin{align}\label{lorentizian_matrix_model}
W[{\rm Rindler}] = \frac{   \int dA   ~e^{ -\frac{8\pi^2M^2N}{\lambda E^2}{\rm Tr}A^2 }~~ {\rm Tr}e^{i  \tau_P  A }  }
  {   \int dA   ~e^{ -\frac{8\pi^2M^2N}{\lambda E^2}{\rm Tr}A^2 } }
 \end{align}
 which can be integrated exactly to give 
 \begin{align}\label{matrixmodel10}
  W[{\rm Rindler}] = ~L^1_{N-1}\left[ \frac{1}{N} \frac{E^2\tau_P^2}{16\pi^2 M^2}\right]\exp\left[- 
 \frac{1}{N} \frac{E^2\tau_P^2}{32\pi^2 M^2}\right]
  \end{align}
  where $L^1_{N-1}$ is the modified Laguerre polynomial given in equation (\ref{laguerre}). 
 The first observation that we can make about this result is that, since, for any finite value of $N$, $L_{N-1}^1(x)$ is a polynomial in its
 argument, for large $\tau_P$, when $N$ is finite, the expression in equation (\ref{matrixmodel10})  decays like a Gaussian.  This is a more severe dependence than
 the linear in $\tau_P$ phase that is expected in the amplitude for a particle.   We must interpret this result
 as telling us that the amplitude is actually zero when $\tau_P$ is taken to be large.  
 
 In retrospect, the amplitude being zero is not surprising, as the W-boson is accelerating and therefore it is expected to
 emit  bremsstrahlung in the way of soft massless vector particles.   Because of infrared singularities, the amplitude for emitting any finite number of vector particles 
 is zero. This is the reason for the damping, and it can also be seen as coming directly from an infrared singularity.   
 In spite of this, we can still extract something from this discussion.  For this, we note that the graphs where  bremsstrahlung external lines are attached are non-planar. 
 These amplitudes are suppressed in the large $N$ limit. This means that the damping would be expected to go away in the large $N$ limit.  (Here we are taking the large $N$ limit before the large $\tau_P$ limit.)  To see that this solves our immediate problem, we take the large $N$  limit  
 of equation (\ref{matrixmodel10}) to get
  \begin{align}\label{matrixmodel11}
  W[{\rm Rindler}] \sim N \frac{ 1}{\sqrt{ \frac{E\tau_P}{4\pi M}}}~J_1\left( \frac{E\tau_P}{2\pi M}\right)
 \end{align}
 where $J_n(x)$ is a Bessel function. 
 Then, we take the large $\tau_P$ limit and we find 
  \begin{align}\label{matrixmodel12}
  W[{\rm Rindler}] \sim N \sqrt{\frac{2}{\pi} }\frac{ 4\pi M}{ { {E\tau_P} }} ~ \exp\left(i \frac{E\tau_P}{2\pi M}\right)
 \end{align}
 where we have used the asmyptotic limit $J_1(x)\sim\sqrt{\frac{2}{\pi x}} \cos(x)$ and we have assumed that an $i\epsilon$ prescription faithfully carried through the 
 above reasoning (it should replace $M$ by $M-i\epsilon$) picks out the appropriate phase.  
In equation (\ref{matrixmodel12}), we have found the oscillating behaviour that is expected for a correction to the propagator.  
Moreover, we can take the coefficient of the linear growth in $\tau_P$ in the phase seriously.  
 Including the contribution of the Wilson loop, the amplitude in the large $N$ limit is then
 \begin{align}\label{punch0}
  {\mathcal A}_{ab}  \approx \delta_{ab}  e^{-i\left(M-\frac{1}{2}M  - \frac{\sqrt{\lambda}E}{2\pi M}\right) \tau_P} 
\end{align}
This result is interesting as it suggests a shift of the rest energy of the accelerated particle by $M\to M-\sqrt{\lambda}\frac{E}{2\pi M}$. 
The Unruh temperature of the acceleration is 
\begin{align}\label{Unruh_temperature}
T_U=\frac{1}{2\pi}\frac{E}{M}
\end{align}
where we have used the fact that the proper acceleration is $\frac{E}{M}$. Then, the energy shift could perhaps be interpreted as
 $M\to M- \sqrt{\lambda} T_U$ which could perhaps be interpreted as a free energy (H=E-TS) with $E=M$, the internal energy, 
  $T$ the Unruh temperature (which coincides with the
 Hawking temperature on the world-sheet) and $S=\sqrt{\lambda}$.

Let us examine this system in the strong coupling limit.  Again, we  must consider the semi-classical limit of the string sigma model which describes
an open string ending on a D3-brane which is suspended at radius $r_M$ in $AdS_5$, and located at a point $\hat \theta_0$ on the 5-sphere.  
The Polyakov action for the bosonic coordinates of the string, in the conformal gauge, is
$$
S=-\frac{L^2}{4\pi \alpha'}\int d\tau d\sigma \eta^{ab}\left\{ \frac{1}{r^2}\partial_a r \partial_b r +r^2 \partial_a x_\mu \partial_b x^\mu + \partial_a \hat\theta
\cdot\partial_b\hat\theta\right\}+\int_{r=r_M}\frac{E}{2}(x_0\dot x_1 - x_1\dot x_0)
$$
where $\eta^{ab}=\left[ \begin{matrix}  -1 & 0 \cr 0 & 1 \cr \end{matrix}\right]$ and $\partial_a=\left[ \begin{matrix} \partial_\tau\cr\partial_\sigma\cr\end{matrix}\right]$.  
This leads to the same equations of motion
$$
-\partial_a\left(\frac{1}{r^2}\partial_a r\right)+r\partial_a x^\mu\partial_a x_\mu-\frac{1}{r^3}\partial_a r\partial_a r=0
$$
$$
\partial_a(r^2\partial_a x_\mu )=0
$$
$$
\partial_a^2\hat\theta-\hat\theta (\hat\theta\cdot\partial_a^2\hat\theta) =0~~,~~\hat\theta^2=1
$$
 and Virasoro constraints 
 $$
  r^2\partial_\sigma x_\mu \partial_\tau x^\mu + \frac{1}{r^2}
 \partial_\sigma r\partial_\tau r +\partial_\sigma\hat\theta\cdot\partial_\tau\hat\theta=0$$
 $$
  r^2 \partial_\sigma x_\mu\partial_\sigma  x^\mu + r^2\partial_\tau x_\mu \partial_\tau x^\mu+\frac{1}{r^2}\partial_\tau r\partial_\tau r+ \frac{1}{r^2}
  \partial_\sigma r \partial_\sigma r+\partial_\sigma\hat\theta\cdot\partial_\sigma\hat\theta + \partial_\tau\hat\theta\cdot\partial_\tau\hat\theta=0
 $$
 that we found before.  We have copied them here for the reader's convenience. In addition, there are boundary conditions, to be imposed at the
 end of the string which intersects the D3 brane, 
\begin{align}\label{bc1}
 \frac{\sqrt{\lambda}}{2\pi }r_M^2\partial_\sigma x_0(\sigma_0,\tau)  +E\partial_\tau x_1(\sigma_0,\tau)=0
 \\ \label{bc2}
  \frac{\sqrt{\lambda}}{2\pi }r_M^2\partial_\sigma x_1(\sigma_0,\tau)  + E\partial_\tau x_0(\sigma_0,\tau)=0
  \\ \label{bc3}
  \frac{\sqrt{\lambda}}{2\pi }r_M^2\partial_\sigma x_{2,3}(\sigma_0,\tau)=0~,~r(\sigma_0,\tau)=r_M
  \end{align}
  It is through these boundary conditions that the information about the presence of the electric field enters the
  dynamics of the string. 
  We shall solve these equations with $x_2=x_3=0$ and $\hat\theta = \hat\theta_0$, a constant. 
 The equations of motion for the remaining variables  have a solution given by the replacement of Euclidean by real time in equation (\ref{embedding}),   
 \begin{align}\label{embedding1}
 x_0(\sigma,\tau) =b\frac{\sinh a\tau}{\cosh a\sigma}
~,~ x_1(\sigma,\tau)=b\frac{\cosh a\tau}{\cosh a\sigma }
~,~ r(\sigma,\tau) =\frac{1}{b} \coth a\sigma 
\end{align}
These are a solution of the equations of motion and the Virasoro constraints.  They must be
adjusted to fit the boundary conditions.  They obey the boundary conditions (\ref{bc1}) and (\ref{bc2}) when
$$
\tanh a\sigma_0 = \frac{2\pi E}{\sqrt{\lambda}r_{M}^2} = \frac{\sqrt{\lambda}}{2\pi}\frac{E}{M^2}
$$
or
$$
\cosh a\sigma_0  = \frac{ 1}{\sqrt{  1-\left( \frac{\sqrt{\lambda}}{2\pi}\frac{E}{M^2}\right)^2}}
~,~
\sinh a\sigma_0  = \frac{ \frac{\sqrt{\lambda}}{2\pi}\frac{E}{M^2}}{\sqrt{  1-\left( \frac{\sqrt{\lambda}}{2\pi}\frac{E}{M^2}\right)^2}}
$$
and the condition (\ref{bc3}) when
$$
 r(\sigma_0) =r_M =  \frac{1}{b}\coth a\sigma_0 =  \frac{1}{b}\frac{\sqrt{\lambda}r_{M}^2}{2\pi E}~\to~b=\frac{M}{E}
 $$
 
  We can also adjust $a$ so that $\tau$ records the proper time of the trajectory, 
\begin{align}\label{acceleration}
  a = \frac{\frac{E}{M}}   { \sqrt{ 1- \left( \frac{\sqrt{\lambda}}{2\pi}\frac{E}{M^2}\right)^2 }}  
  \end{align}
  Here, the worldsheet coordinates lie in the ranges $-\tau_P/2<\tau< \tau_P/2$ and $\sigma_0\leq\sigma<\infty$. 
 
The solution that we have found is  the locus of the curve
$$
(x_0)^2-(x_1)^2-\frac{1}{r^2}+\frac{M^2}{E^2}=0
$$
which, at radius $r=r_M$ the endpoint of the string traces a Rindler trajectory, but with   acceleration given in equation (\ref{acceleration})
which is enhanced from what
it would be for a particle with the same charge and mass.
This acceleration diverges at what was argued in reference \cite{Semenoff:2011ng} to be an upper critical electric field.  
In fact, remembering that $M=\frac{\sqrt{\lambda}r_M}{2\pi}$  we can see that this is the same upper critical field that we
would deduce from the Born-Infeld action in equation (\ref{d3_dbi}).   This means that, when the electric field is strong enough
that the combination   $ \frac{\sqrt{\lambda}}{2\pi}\frac{E}{M^2}$ is of order one, we expect that Schwinger pair production becomes important
and   that it competes with the process that we are considering.  Since we did not take it into account in our quantum field theory
computation, the best that we can do for comparison with that computation is to assume that $ \frac{\sqrt{\lambda}}{2\pi}\frac{E}{M^2}<<1$.
In the following, we shall make this assumption by keeping only terms which are linear in $ \frac{\sqrt{\lambda}}{2\pi}\frac{E}{M^2}$. 

The induced metric of the world-sheet is 
\begin{align}\label{metric_accelerated}
 ds^2 =L^2\left(\frac{a}{\sinh a\sigma}\right)^2 \left[ d\sigma^2-d\tau^2   \right]
\end{align}
  which is just the analytic continuation of the Euclidean expression that we found in the context of the circle Wilson loop (\ref{ads2blackhole}) where
  the parameter $a$ is now the acceleration.  This is a Lorentzian signature $AdS_2$ black hole metric. The general solution for the world-sheet of a
  string with any time-like boundary placed at the boundary of $AdS_5$ was found by Mikhailov \cite{mikhailov}.  Our solution agrees with it in the 
  appropriate limit. 
  
    It is very interesting that the  Hawking temperature of this classical world-sheet, which is given in equation (\ref{Hawking_temperature}),
  coincides with the Unruh temperature of the space-time acceleration which was given in equation (\ref{Unruh_temperature})
  $$
  T_H = \frac{a}{2\pi} =\frac{1}{2\pi}  \frac{\frac{E}{M}}   { \sqrt{ 1- \left( \frac{\sqrt{\lambda}}{2\pi}\frac{E}{M^2}\right)^2 }} =T_U
  $$
  It is also interesting that both of these temperatures diverge, like the acceleration does, at the critical electric field. 
  
  The event horizon of the metric (\ref{metric_accelerated}) is located at $\sigma\to\infty$ which is at $r=\frac{E}{M}$.  
  If we compute the action that is due only to the worldsheet which is located above the event horizon, we obtain
  \begin{align}
  S=\left( 1    -\frac{\sqrt{\lambda}}{2\pi}  \frac{E}{M^2}
 \right)   M \tau_P    -\frac{M}{2}\tau_P +\ldots  \label{punch1}
 \end{align}
 and the string theory prediction for the amplitude is 
 \begin{align}
 {\mathcal A}_{ab} = ~{ \delta_{ab}}~ e^{iM\tau_P-i\frac{1}{2}M\tau_P    -i \frac{\sqrt{\lambda}}{2\pi}  \frac{E}{M}\tau_P+\ldots }
  \label{punch2}
  \end{align}
  where the ellipses indicate that, following our discussion above,  we have ignored terms of higher than linear order in $ \frac{\sqrt{\lambda}}{2\pi}\frac{E}{M^2}$. 
 This agrees precisely with our quantum field theory result in equation (\ref{punch0}).   
  
  This rather remarkable agreement suggests that the correct result from the string theory side is, as we have done here, to take the area of the world-sheet which is above the event horizon.  This brings up a paradox as the world-sheet does not just end at the horizon, it can be continued smoothly beyond the horizon and, in fact, as pointed out in reference \cite{Hubeny:2014kma}, it curls back and reaches the boundary region of $AdS_5$ at a second Rindler trajectory for the W-boson's anti-particle.  So, why not the full world-sheet?  The answer could lie  in causality.  As the endpoint of the open string moves with a constant acceleration
  the rest of the string should lag behind it.  As we have seen, this string should sweep out the section of $AdS_2$ which is above the event horizon.   However, the string, does not have time to make the connection through the wormhole to the other parts of the extended would-be world-sheet.  Instead, as advocated in reference \cite{garcia},  a shock-front forms at the event horizon.  The world-sheet is not analytic there, the event horizon is a line where the exterior $AdS_2$ is joined to 
a null surface which descends from the event horizon to the Poincare horizon of $AdS_5$.  Being null, the latter surface has zero proper area and it does not contribute to the on-shell action.

  \section{Epilogue}
  
  In these lectures I have attempted to give a simple introduction to the idea of duality between gauge fields and string theory in its best understood form.  I have made no attempt to be complete or even to address what some would regard as the most important topics.  To my mind, there is little need for that. There is already a lot of excellent literature about this  subject including comprehensive introductory textbooks which I encourage you to look at.  What I hope that I have accomplished is to give some appreciation of the beauty of the ideas which led us to this point.

\end{document}